\documentclass[twocolumn,amsmath,amssymb,floatfix,prb,showpacs,footinbib,superscriptaddress]{revtex4}
\usepackage{amsmath,amssymb,natbib,bm,graphicx,url,psfrag}
\usepackage[ansinew]{inputenc}

\newcommand{\be}{\begin{equation}}
\newcommand{\ee}{\end{equation}}
\newcommand{\bes}{\begin{equation}\begin{split}}
\newcommand{\ees}{\end{split}\end{equation}}

\newcommand{\vc}[1]{\mathbf{#1}}

\newcommand{\abs}[1]{\left|#1\right|}

\newcommand{\ket}[1]{\left|\, #1 \, \right\rangle}

\newcommand{\boket}[3]{\left\langle\, #1 \,|\, #2 \,|\, #3 \,\right\rangle}

\DeclareMathOperator{\tr}{tr}

\DeclareMathOperator{\rre}{Re}
\DeclareMathOperator{\iim}{Im}
\DeclareMathOperator{\sgn}{sgn}
\DeclareMathOperator{\Ll}{L}

\begin{document}
\title{Theory of the Franck-Condon blockade regime}
\author{Jens Koch}
\affiliation{Institut f\"ur Theoretische Physik, Freie Universit\"at Berlin, Arnimallee 14, 14195 Berlin, Germany}
\author{Felix \surname{von Oppen}}
\affiliation{Institut f\"ur Theoretische Physik, Freie Universit\"at Berlin, Arnimallee 14, 14195 Berlin, Germany}
\author{A.~V.~Andreev}
\affiliation{Department of Physics, University of Washington, Seattle, Washington 98195-1560, USA}
\date{November 30, 2006}
\begin{abstract}
Strong coupling of electronic and vibrational degrees of freedom entails a low-bias suppression of the current through single-molecule devices, termed Franck-Condon blockade. In the limit of slow vibrational relaxation, transport in the Franck-Condon-blockade regime proceeds via avalanches of large numbers of electrons, which are interrupted by long waiting times without electron transfer. The avalanches consist of smaller avalanches, leading to a self-similar hierarchy which terminates once the number of transferred electrons per avalanche becomes of the order of unity.  
Experimental signatures of self-similar avalanche transport are strongly enhanced current (shot) noise, as expressed by giant Fano factors, and a power-law noise spectrum. We develop a theory of the Franck-Condon-blockade regime with particular emphasis on effects of electron cotunneling through highly excited vibrational states. As opposed to the exponential suppression of sequential tunneling rates for low-lying vibrational states, cotunneling rates suffer only a power-law suppression. This leads to a regime where cotunneling dominates the current for any gate voltage. Including cotunneling within a rate-equation approach to transport, we find that both the Franck-Condon blockade and self-similar avalanche transport remain intact in this regime. We predict that cotunneling leads to absorption-induced vibrational sidebands in the Coulomb-blockaded regime as well as intrinsic telegraph noise near the charge degeneracy point. 
\end{abstract}
\maketitle
\section{Introduction\label{sec:intro}}
According to many recent experiments,\cite{park4, natelson}  transport through single-molecule junctions is characterized by Coulomb-blockade behavior and exhibits a Kondo effect at low temperatures, in close analogy with experiments on quantum dots. An important distinction between transport through quantum dots and single molecules lies in the coupling to phonon degrees of freedom. While charge carriers typically interact with a continuum of phonon modes in quantum dots, molecules are characterized by a discrete spectrum of vibrational modes. Recent experiments on single-molecule junctions have shown that these discrete excitations are reflected in current-voltage characteristics ($IV$s)  as vibrational sidebands.\cite{park,zhitenev,ho,chae} This experimental effort has motivated much theoretical work.\cite{schoeller3,varma,flensb1,aleiner} It is now understood that transport through single-molecule junctions exhibits a number of qualitatively different regimes, depending on parameters such as the strength of the electron-phonon coupling and the vibrational relaxation rate, see e.g.\ Refs.\ \onlinecite{koch2,wegewijs1,koch6,ryndyk}.

The importance of Coulomb-blockade physics in single-molecule junctions suggests that molecule and leads are often weakly coupled from the point of view of electronic transport. Hence, perturbation theory in the tunneling amplitude between molecule and leads should be good starting point for a theoretical description. 
Such a perturbative treatment is valid if the tunneling-induced level width $\Gamma$ is small compared to the temperature $T$. Then, the lowest order in this expansion results in \emph{sequential tunneling}, which corresponds to the transfer of one electron from a lead onto the molecule or vice versa. Sequential tunneling is dominant as long as the electronic level is situated between the Fermi energies of the leads. By contrast, if the level is located outside of the bias window, sequential tunneling is exponentially suppressed, and (next-to-leading order) \emph{cotunneling} processes take over.\cite{averin2} This marks the onset of Coulomb blockade.\cite{averin4} In many experiments, a gate electrode can be used to tune the system between the sequential tunneling and the cotunneling regime.

The coupling of electron tunneling to molecular vibrations adds a wealth of new physics. This coupling originates from the fact that the nuclear configuration of the molecule needs to adjust to the addition of electrons to the molecule. Thus, electron transfer is generally accompanied by excitation (or deexcitation) of molecular vibrations. In the simplest case, the electron-phonon coupling can be modeled by a charge-dependent displacement of the molecular potential surface.

In two recent papers,\cite{koch2,koch3} two of us have investigated the regime of strong electron-phonon coupling, where the displacements of the potential surfaces are large compared to the quantum fluctuations of the nuclear configuration in the vibrational ground state. As a result, the overlap between low-lying vibrational states is drastically suppressed. It is this suppression of the Franck-Condon matrix elements which leads to a strong low-bias suppression of the sequential-tunneling current, which we termed \emph{Franck-Condon (FC) blockade}. It is a hallmark of the FC  blockade that it cannot be lifted by means of a gate voltage, in striking contrast with other blockades such as the Coulomb blockade.

It is evident that the FC blockade is more severe for molecules with fast vibrational relaxation (arising, e.g.. due to interactions with the substrate, radiation, or coupling to other vibrational modes). The reason is that for slow vibrational relaxation, the transport current excites the molecular vibrations, and transport involves excited vibrational states for which the suppression of the FC matrix elements is less pronounced. Nevertheless, the Franck-Condon blockade in the current-voltage characteristics exists irrespective of the strength of vibrational relaxation.

It is thus remarkable that the transport mechanisms in the limits of weak and strong relaxation are in fact fundamentally different, as revealed by current shot noise. In the sequential-tunneling approximation, strong relaxation leads to purely sub-Poissonian current noise, as commonly expected for transport in fermionic systems.\cite{blanter} Electrons are transferred one by one, slightly avoiding each other due to the Pauli principle. By contrast, for weak vibrational relaxation transport is dominated by \emph{self-similar avalanches of electrons} with giant Fano factors reflecting the large  number of electrons per avalanche. The occurence of avalanches is explained by the exponential growth of sequential-tunneling rates with increasing vibrational quantum number. Whenever the molecule is in its vibrational ground state, the suppressed FC matrix elements require a long waiting time before an electron tunnels. Most likely, the transition involves an excitation of the vibrations. Now, having left the vibrational ground state, the accessible transitions have faster rates, leading to rapid tunneling dynamics and avalanche formation. If, eventually, the system undergoes a transition back into the vibrational ground state, the avalanche terminates and another long waiting time is induced. The self-similarity arises, since all previous arguments can also be applied to the first excited vibrational state. As an experimental fingerprint of the self-similarity, we have predicted a characteristic power-law decay of the current noise as a function of frequency.

\begin{figure}
	\centering
		\includegraphics[width=0.8\columnwidth]{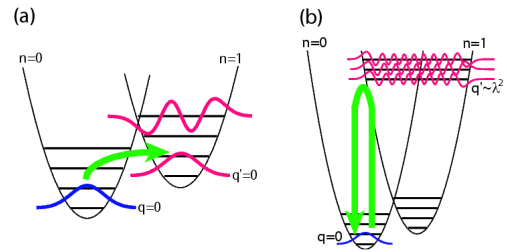}
			\caption{(Color online) (a) Sequential tunneling versus (b) cotunneling for strong electron-phonon coupling. In the FC blockade regime, sequential processes as depicted in (a) are strongly suppressed due to the exponentially small overlap of harmonic oscillator wavefunctions. For cotunneling processes (b), this suppression is partially lifted due to contributions from highly excited virtual phonon states.
			\label{fig-seqcot}}
\end{figure}

In this paper, we investigate the cotunneling contributions to current and noise in the FC blockade regime. The inclusion of such higher-order processes is motivated by the insight that sequential-tunneling rates are exponentially suppressed in the electron-phonon coupling. By contrast, cotunneling may benefit from the participation of highly excited virtual phonon states with much better overlap, see Fig.~\ref{fig-seqcot}. In this way, the relevant FC matrix elements are no longer exponentially suppressed, which may render cotunneling relevant despite its suppression in powers of the small quantity $\Gamma/k_BT$.
The central question which we address in this paper is whether and if so, in which way, cotunneling affects the phenomenology of the Franck-Condon blocakade as sketched above. We investigate this question within the rate-equations approach, extending the formalism to include cotunneling corrections. 

For the convenience of readers who are less interested in the technical details, we begin our presentation in Section \ref{sec:qual-disc} with a summary of our principal results. 
In Section \ref{sec:model} we present the model that our analysis is based upon. The computational formalism and the question of regularized cotunneling rates are addressed in Section \ref{sec:formalism}. In particular, we discuss the cotunneling rates in the FC blockade regime, and determine their scaling with the electron-phonon coupling. Section \ref{sec:current} contains the investigation of cotunneling corrections to the linear conductance and nonlinear current-voltage characteristics. Further insight into the transport mechanism and the fate of avalanches is provided in Section \ref{sec:noise}, where we discuss the current shot noise beyond the sequential-tunneling approximation.
Finally, the conclusions from this analysis are summarized in Section \ref{sec:conclusions}. Technical details concerning the calculations are deferred to appendices.

%

\section{Qualitative discussion\label{sec:qual-disc}}
\subsection{The fate of the FC blockade}
The increased overlap of vibrational wavefunctions exploited by cotunneling processes indeed makes cotunneling corrections significant in the FC blockade regime. In contrast to sequential-tunneling rates, we find that cotunneling rates are not exponentially, but rather algebraically suppressed in the electron-phonon coupling, see Eq.~\eqref{cotappr}. Consequently, for sufficiently strong electron-phonon coupling, cotunneling dominates the linear conductance over the entire gate-voltage range, see Fig.~\ref{fig-lincond}. In particular, the paradigm of a gate-controlled crossover between sequential tunneling and cotunneling breaks down in this regime, and the linear conductance develops a kink at the degeneracy point. Cotunneling also significantly increases the low-bias current well beyond the linear-response regime, see Fig.~\ref{fig1}. More importantly however, we find that the remaining strong algebraic suppression of cotunneling rates  still leads to a current reduction. Therefore, a central finding is that the FC blockade as such \emph{remains intact}.

\subsection{Persistence of avalanches and enhanced Fano factors}
Another central result is that the avalanche dynamics, found in the sequential-tunneling approximation for weak vibrational relaxation, \emph{persists} in the presence of cotunneling. This can be understood by noting that cotunneling is dominant, whenever the molecule is close to its vibrational ground state. However, the avalanches involve excited vibrational states for which cotunneling is in fact a small perturbation. Nevertheless, the long waiting times between
avalanches are now governed by cotunneling, leading to Fano factors which are smaller than predicted by the sequential-tunneling approximation, although still strongly enhanced compared to the Poissonian limit, see Fig.~\ref{fig-fanoun}. In a similar vein, the power-law behavior of the noise power spectrum remains valid, but is reduced in the frequency range over which it applies, see Fig.~\ref{fig-spectrum}.

\subsection{Absorption-induced vibrational sidebands}
For weak vibrational relaxation, cotunneling leads to a striking new feature: the appearance of additional vibrational sidebands \emph{inside} the Coulomb-blockaded region, see Fig.~\ref{fig2}. Inside the Coulomb-blockaded regime, molecular vibrations can be excited by inelastic cotunneling. 
Once molecular vibrations are excited, sequential tunneling processes become possible when accompanied by absoption of a phonon. This leads to new vibrational sidebands which were in fact observed in experiments on suspended carbon nanotubes.\cite{LeRoy} Interestingly, such absorption-induced sidebands are expected to be a generic feature which is not limited to strong electron-phonon coupling.

\subsection{Inherent telegraph noise}
\begin{figure}
	\centering
		\includegraphics[width=0.75\columnwidth]{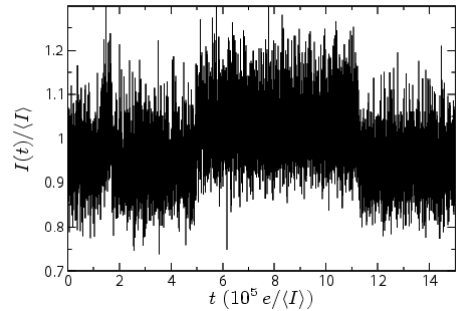}
			\caption{Coarse-grained current $I(t)$ as obtained by Monte-Carlo simulation for equilibrated phonons at strong electron-phonon coupling ($\lambda=5$), with bias and gate voltage fixed to $eV=0.8\hbar\omega$ and $\epsilon_d=0.5\hbar\omega$, respectively.  The time-dependence of the current clearly reveals the current telegraph noise caused by a slow switching between the two molecular charge states via sequential tunneling. ($k_BT=0.05\hbar\omega$, $\Gamma_a=0.02\hbar\omega$; coarse graining over 300 events)
			\label{fig-telegraph}}
\end{figure}

For strong vibrational relaxation, avalanche dynamics is ruled out and  current noise remains sub-Poissonian in the sequential-tunneling approximation. Surprisingly, we find that cotunneling causes a strong enhancement of current noise at low bias due to inherent telegraph noise, see Figs.~\ref{fig-telegraph} and \ref{fig-fanoeq}. Traces of this telegraph behavior are also found for weak vibrational relaxation, rendering this effect robust with respect to the relaxation rate of the system. Its origin is the unusual situation that cotunneling rates are larger than  sequential-tunneling rates, $W_\text{cot}\gg W_\text{seq}$. Thus, sequential tunneling leads to a slow switching between the two charge states $n=0$ and $1$, which modulates the fast cotunneling dynamics. The cotunneling rates in these two states are different in general, which leads to two distinct current states. This is reflected in large Fano factors $F\sim \Delta W_\text{cot}/W_\text{seq}\gg1$, where $\Delta W_\text{cot}$ denotes the difference of cotunneling rates in the two charge states.

\section{Model\label{sec:model}}
The Anderson-Holstein Hamiltonian
constitutes the minimal model which captures Coulomb blockade, the Kondo effect, as well as vibrational sidebands.\cite{glazman2,wingreen,flensb1,aleiner,koch2,paaske,wegewijs1,wegewijs2} Several assumptions and simplifications enter into this model: (i) Transport is assumed to be caused by tunneling through one spin-degenerate orbital of the molecule. (ii) One mode of molecular vibrations is taken into account within the harmonic approximation. These ingredients result in the Hamiltonian $H= H_\text{mol} + H_\text{leads} + H_\text{T}$ with
\begin{align}
H_\text{mol}= &\epsilon_d n_d + Un_{d\uparrow}n_{d\downarrow} + \hbar\omega b^\dag b  + \lambda \hbar\omega (b^\dag + b)n_d,  \label{Hmol}
\end{align}
describing the electronic and vibrational degrees of freedom of the molecule,
\be
H_\text{leads}= \sum_{a=L,R}\sum_{\vc{p},\sigma} (\epsilon_{a\vc{p}}-\mu_a)c^\dag_{a\vc{p}\sigma}c_{a\vc{p}\sigma}
\label{Hleads}
\ee
the noninteracting leads, and 
\be
H_\text{T}= \sum_{a=L,R}\sum_{\vc{p},\sigma} \left( t_{a\vc{p}} c^\dag_{a\vc{p}\sigma} d_\sigma + \text{h.c.}\right)
\label{Hmix}
\ee
the tunneling between leads and molecule. (iii) In addition, electronic relaxation\index{electronic relaxation} is taken to provide the shortest time scale in the problem, so that the metallic electrodes are described by Fermi seas in thermal equilibrium at all times. 

The validity  and various extensions of this model in the context of transport through single molecules have been discussed, e.g., in Refs.~\onlinecite{flensb1,aleiner,koch4,koch5,wegewijs3,wegewijs1}. However, the inclusion of generalizations would only obscure the central points of the present paper, and we therefore restrict ourselves to the model constituted by Eqs.~\eqref{Hmol}-\eqref{Hmix}. For simplicity, we consider the case of symmetric voltage splitting, i.e.\ the bias voltage $V$ enters the left and right Fermi energies as $\mu_{L,R}= \pm eV/2$. (Note that all energies are measured with respect to the zero-bias Fermi energy.) All results are derived for the wide-band limit with constant densities of states $\rho_{L,R}$ and momentum-independent tunneling matrix elements $t_a$.

There are four different energy scales relevant to the analysis of transport regimes of the Anderson-Holstein model: the phonon energy $\hbar\omega$, the tunneling-induced level width $\Gamma=2\pi\sum_a\rho_a\abs{t_a}^2$, and the energy scales set by the Kondo temperature $T_K$ and the ambient temperature $T$. Here, we are interested in the ``high-temperature" limit of the model, i.e.\ we study the regime where $k_BT_K,\,\Gamma \ll \hbar\omega,\,k_BT$. In this case, the tunneling Hamiltonian $H_\text{T}$ may be treated perturbatively, and it has been shown that the equation of motion for the reduced density matrix of the molecule reduces to simple rate equations.\cite{aleiner} 

The electron-phonon coupling, parametrized by the dimensionless coupling parameter $\lambda$, can be treated exactly by means of the Lang-Firsov canonical transformation.\cite{lang} This transformation amounts to renormalizations of the single-particle energy $\epsilon_d\rightarrow\epsilon_d' = \epsilon_d - \lambda^2\hbar\omega$, and of the charging energy $U\rightarrow U'=U-2\lambda^2\hbar\omega$  (polaron shift), and it introduces a translation operator into the tunneling matrix elements, $t_a\to \exp[-\lambda(b^\dag-b)]t_a$. This shift $\sim\lambda$ in the normal coordinate of the harmonic oscillator precisely corresponds to the fact that the equilibrium distances between the nuclei of the molecule change whenever an additional electron is present in the molecule. The transformed Hamiltonian reads $H'=H_\text{mol}'+H_\text{leads}'+H_\text{T}'$ with $H_\text{leads}'=H_\text{leads}$, and
\begin{align}
H_\text{mol}'= &\epsilon_d' n_d + U'n_{d\uparrow}n_{d\downarrow} + \hbar\omega b^\dag b, \\
H_\text{T}'=& \sum_{a=L,R}\sum_{\vc{p},\sigma} \left( t_{a\vc{p}}e^{-\lambda(b^\dag-b)} c^\dag_{a\vc{p}\sigma} d_\sigma + \text{h.c.}\right).
\end{align}
In the following, we will proceed with the transformed Hamiltonian and drop all primes.

The system is expected to display qualitatively different behavior depending on the vibrational relaxation rate as well as the strength of the electron-phonon coupling. 
Throughout this paper, we will focus on the case of strong electron-phonon coupling, $\lambda\gg1$,\footnote{Note that the overlap of harmonic oscillator wave functions decreases \emph{exponentially} with the coupling strength $\lambda$. Consequently, strong electron-phonon coupling is quickly reached for values $\lambda \agt 3$.} and nonnegative effective charging energies, $U-2\lambda^2\hbar\omega\ge 0$. The latter inequality is always satisfied for sufficiently strong Coulomb repulsion between electrons inside the molecular orbital. Unless explicitly stated otherwise, we will concentrate on the limit $U\to\infty$, applicable whenever $U$ is large compared to all other relevant energy scales.

\section{Formalism\label{sec:formalism}}
In this section, we discuss the formalism which constitutes the basis for our analysis. The main input for the transport calculations consists of the transition rates induced by electron tunneling. In the following subsection, we discuss how to obtain these rates, especially in the case of cotunneling where the naive application of Fermi's golden rule leads to a divergence. The subsequent subsection explains the rate-equations approach for calculating the steady-state current as well as the noise spectrum.

\subsection{Transition rates}
Regarding the tunneling Hamiltonian $H_T$  as a perturbation, one can proceed to calculate rates for transitions $\ket{i}\to\ket{f}$ by an expansion of the $T$-matrix and applying Fermi's golden rule,
\begin{align}\label{goldenrule}
\gamma_{if}&=\frac{2\pi}{\hbar} \abs{\boket{f}{T}{i}}^2 \delta(E_i-E_f)\\\nonumber
&= \frac{2\pi}{\hbar} \abs{\boket{f}{H_T+H_TG_0H_T+\cdots}{i}}^2\delta(E_i-E_f).
\end{align}
Here, $G_0=[E_i-H_\text{mol}-H_\text{leads}+i\eta]^{-1}$ is the free retarded Green's function. Generally, the initial and final states $\ket{i}$, $\ket{f}$ still involve degrees of freedom of the leads. These are eliminated by integrating over the energies of particles and holes generated in the Fermi seas during the process, taking into account thermal occupations by $f$ or $1-f$ factors, where $f$ is the Fermi function. As a result, one arrives at rates $W$ which are labelled by molecular degrees of freedom only. (Of course, the rates still include a label for the participating junction(s).)

To lowest order in $H_T$, one obtains \emph{sequential tunneling} processes, which transfer one electron from a lead onto the molecule or vice versa. For example, the rate for a process changing the charge number of the molecule from $n=0$ to $n=1$ by tunneling across junction $a$, and simultaneously changing the number of excited phonons from $q$ to $q'$, is given by
\be
W^{01}_{qq';a}=s(0,1)\frac{\Gamma_a}{\hbar}\abs{M_{qq'}}^2f_a(\epsilon_d + [q'-q]\hbar\omega).
\label{seqrate1}
\ee
Here, $f_a(\epsilon)=f(\epsilon-\mu_a)$ denotes the Fermi function for lead $a$. The symbol $M_{qq'}$ denotes the FC matrix elements for a phonon transition $q\to q'$, given by the overlap of two harmonic oscillator wavefunctions $\phi$, spatially displaced by the distance $\Delta x=\sqrt{2}\lambda\ell_\text{osc}$. Here, $\ell_\text{osc}=\sqrt{\langle x^2 \rangle_{q=0}}=(\hbar/M\omega)^{1/2}$ defines the oscillator length. Introducing $q=\min\{q_1,q_2\}$ and $Q=\max\{q_1,q_2\}$, the matrix elements read
\begin{align}\label{FCmat}
M_{q_1q_2}=&\int_{-\infty}^\infty dx\, \phi_{q_1}^*(x+\Delta x)\phi_{q_2}(x)\\\nonumber =&\left[\sgn(q_2-q_1)\right]^{q_1-q_2}\lambda^{Q-q}e^{-\lambda^2/2}\sqrt{\frac{q!}{Q!}}\Ll_{q}^{Q-q}(\lambda^2),
\end{align}
where $\Ll_m^{n}(x)$ denotes the generalized Laguerre polynomial.

Our choice of jointly treating the spin-up and the spin-down states as the singly-occupied state $n=1$ requires the inclusion of a spin factor $s$.  The definitions $s(0,1)\equiv2$ and $s(1,0)\equiv1$ reflect the fact that rates for processes $n=0\to 1$ are twice as large as rates for $n=1\to 0$ due to the spin-degeneracy of the state $n=1$. The rate for the analogous process $n=0\to1$ can be obtained from Eq.~\eqref{seqrate1} by swapping the spin factor and substituting $f_a$ by $(1-f_a)$, i.e.\
\be
W^{10}_{qq';a}=s(1,0)\frac{\Gamma_a}{\hbar}\abs{M_{qq'}}^2\left\{1-f_a(\epsilon_d - [q'-q]\hbar\omega)\right\}.
\label{seqrate2}
\ee

\begin{figure}
	\centering
		\includegraphics[width=0.8\columnwidth]{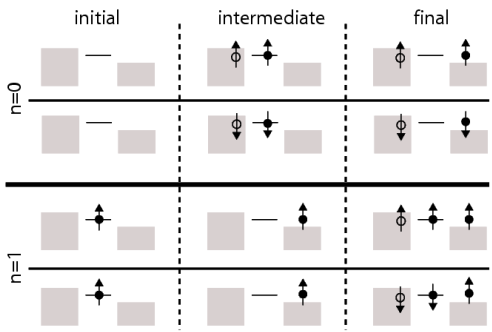}
			\caption{Relevant contributions to the cotunneling rates in the $U\to\infty$ limit. For both charge states $n=0$ and $1$ there are two incoherent contributions with either a spin-up or a spin-down electron being transferred across the molecule.
			\label{fig-cotprocesses}}
\end{figure}

In next-to-leading order in $H_T$, \emph{cotunneling} processes are generated. These transfer one electron from lead $a$ to lead $b$, while the electronic occupation of the molecule changes only virtually in the intermediate state. Based on Eq.~\eqref{goldenrule}, the corresponding rates are obtained as
\begin{widetext}
\begin{align}
\label{cotrate1}
W^{00}_{qq';ab}&=\frac{s(0,0)}{2\pi\hbar}\Gamma_a\Gamma_b \int d\epsilon\, \abs{\sum_{q''}\frac{M_{q'q''}M_{qq''}^*}{\epsilon-\epsilon_d+(q-q'')\hbar\omega}}^2f_a(\epsilon)\left[1-f_b(\epsilon+[q-q']\hbar\omega)\right]\\
W^{11}_{qq';ab}&=\frac{s(1,1)}{2\pi\hbar}\Gamma_a\Gamma_b \int d\epsilon\, \abs{\sum_{q''}\frac{M_{q'q''}M_{qq''}^*}{\epsilon_d-\epsilon+(q'-q'')\hbar\omega}}^2f_a(\epsilon)\left[1-f_b(\epsilon+[q-q']\hbar\omega)\right].
\label{cotrate2}
\end{align}
\end{widetext}
Here, the spin factors are $s(0,0)=s(1,1)\equiv 2$, which follows from an analysis of the relevant contributions as illustrated in Fig.~\ref{fig-cotprocesses}. For cotunneling through the empty molecule, $n=0\to0$, there is an incoherent addition of the processes transferring a spin-up or a spin-down electron, respectively. For $n=1\to1$ the cotunneling transition may either leave the molecule's spin state invariant or cause a spin flip, again resulting in two incoherent processes. In both cases $n=0$ and $1$, the rates of the two contributions are identical and hence can be absorbed into the spin factor.

It is crucial to note that, in general, the cotunneling rates cannot be directly evaluated from Eqs.~\eqref{cotrate1} and \eqref{cotrate2}. These expressions diverge due to second-order poles from the energy denominators. This divergence, related to the infinite lifetime of the virtual intermediate state within a purely perturbative approach, has been pointed out before.\cite{averin,turek} A regularization scheme has been developed in order to extract the correct cotunneling rates,\cite{turek,koch} which we summarize here, deferring the details to Appendix \ref{app:cotrates}.

The basis for the regularization of cotunneling rates is given by two observations. (i) Second-order perturbation theory clearly misses the fact that the intermediate state obtains a finite width $\sim\Gamma$ due to the tunneling. This width should enter the energy denominator as an imaginary part, shifting the pole away from the real axis. (ii) For a specific cotunneling transition $\ket{i}\to\ket{f}$ at finite temperature, the final state $\ket{f}$ can also be reached from the initial state $\ket{i}$ by two sequential processes.


Specifically, the regularization proceeds as follows. First, a level width $\gamma\sim\Gamma$ is introduced in the energy denominators. With the poles shifted away from the real axis, the integral can indeed be carried out. The resulting expression is then expanded in powers of the level width $\gamma$. The leading order term is proportional to $1/\gamma$. When combining this with the prefactor $\Gamma_a\Gamma_b$ of the rates Eqs.~\eqref{cotrate1} and \eqref{cotrate2}, the overall order of this term is found to be identical to a sequential-tunneling term. Following point (ii), this term must be disregarded to avoid double-counting of sequential processes. The next-to-leading order in the $\gamma$-expansion is a $\gamma^0$ term. This term gives the regularized expression for the cotunneling rate. The full regularization procedure is carried out in detail in Appendix \ref{app:cotrates}, and the resulting rates are given in  Eqs.~\eqref{reg1} and \eqref{reg2}. 

It is interesting to note that these ``cotunneling rates" may become \emph{negative}. This result can in fact be expected by considering the simple case of a single-level resonant tunneling model without charging energy. For this model, the exact result for the current in the $T\to0$ limit is  
\be
I=\sgn(eV)\frac{2e}{h}\frac{\Gamma_L\Gamma_R}{\Gamma_L+\Gamma_R}\sum_{a=L,R} a\arctan\left[\frac{a\abs{eV}-2\epsilon_d}{\Gamma}\right],
\ee
where we identify the index $L$ ($R$) with $a=+1$ ($a=-1$). Expanding this in powers of $\Gamma$, we find that the second-order term corresponding to cotunneling is given by
\be
I_\text{cot}=-\frac{e}{h}\Gamma_L\Gamma_R\frac{\abs{eV}}{(eV/2)^2-\epsilon_d^2}.
\ee
This expression indeed becomes negative, whenever the level is located inside the bias window, i.e.\ cotunneling \emph{reduces} the sequential tunneling. Our regularization procedure combined with the rate-equations approach exactly reproduces these expressions, as shown in detail in Appendix \ref{app:exact}. It is worth noting that, alternatively, the formalism developed by K{\"o}nig, Schoeller and Sch{\"o}n may be used to circumvent the necessity of regularization, see e.g.\ Ref.~\onlinecite{koenigPRL}. We find that additional corrections due to level shifts and broadening captured by their approach are irrelevant in the case $\Gamma\ll k_BT$ considered here. In this limit, both approaches are expected to give identical results.

In the FC blockade regime, the dominant contributions to the cotunneling rates [Eqs.\ \eqref{cotrate1}, \eqref{cotrate2}, and \eqref{reg1}, \eqref{reg2}] arise from highly excited virtual phonon states. According to the Franck-Condon principle, the wavefunction overlap reaches a maximum for transitions that occur (approximately) vertically in the diagram Fig.\ \ref{fig-seqcot}(b). For transitions originating in the vibrational ground state $q=0$, the ``optimal" overlap is obtained for virtual states $q''$ of the order of $q''\approx\lambda^2$. Deep inside the FC blockade, the resulting poles $\epsilon\approx \epsilon_d \pm \lambda^2\hbar\omega$, see Eqs.\ \eqref{cotrate1}, \eqref{cotrate2}, are typically located well beyond the bias window. As a consequence of that, negative cotunneling rates are essentially irrelevant in the FC blockade regime.\footnote{This statement strictly holds for the equilibrated case $\tau\to0$. For the unequilibrated case $\tau\to\infty$, the dynamics within avalanches involving highly excited phonon states may acquire small negative corrections due to cotunneling.}

For an understanding of cotunneling rates  it is important to note that there is not a single virtual state $q''$ resulting in maximal overlap, but rather an extended range $\Delta q\sim\lambda$ of excited states centered around $q''\sim\lambda^2$, see Fig.~\ref{fig-seqcot}(b). By applying Sterling's formula to the expression \eqref{FCmat} for the FC matrix element $M_{0\lfloor\lambda^2\rfloor}$, we obtain $\abs{M_{0\lfloor\lambda^2\rfloor}}^4\approx1/2\pi\lambda^{2}$, so that the contribution from the intermediate state $q''=\lfloor \lambda^2\hbar\omega\rfloor$ is given by $\frac{\Gamma_a\Gamma_b}{\pi\hbar}\frac{k_BT + eV}{\lambda^2(\epsilon_d+\lambda^2\hbar\omega)^2}$. Carrying out the coherent sum over the effective range $\Delta q\sim\lambda$, and subsequent squaring leads to an additional factor of $\lambda^2$, resulting in the approximation 
\be\label{cotappr}
W^{00}_{00;LR}\approx \frac{\Gamma_L\Gamma_R}{\pi\hbar}\frac{k_BT+eV}{(\epsilon_d+\lambda^2\hbar\omega)^2}
\ee
valid for $0\le eV\ll\lambda^2\hbar\omega$. A more detailed derivation of this expression is given in Appendix \ref{app:approx}. By contrast to the sequential-tunneling rates, Eq.~\eqref{cotappr} establishes that cotunneling rates are not exponentially but rather algebraically suppressed. In particular, for $\epsilon_d=0$, we have $W^{00}_{00;LR}\sim\lambda^{-4}$.

\subsection{Current and shot noise}
As shown, e.g., in Refs.~\onlinecite{aleiner} and \onlinecite{korotkov}, the steady-state current and the current noise can be computed via the rate-equations formalism. In the first step, the steady-state occupation probabilities $P^n_q$ are obtained from the rate equations
\begin{align}\label{rateeqs1}
0=\frac{dP^n_q}{dt} =& \sum_{n',q'}\left[ P^{n'}_{q'}W^{n'n}_{q'q} - P^n_q W^{nn'}_{qq'}\right]\\\nonumber
&\quad-\frac{1}{\tau}\left[{\textstyle P^n_q- P^\text{eq}_q \sum_{q'} P^n_{q'}}\right].
\end{align}
The last term is added on a phenomenological basis, and describes relaxation of the vibrations towards the equilibrium distribution $P^\text{eq}_q=e^{-q\hbar\omega/k_BT}(1-e^{-\hbar\omega/k_BT})$ on a time scale $\tau$. In the following, we will mainly investigate the two limiting cases of equilibrated phonons ($\tau\to0$, distribution fixed to equilibrium), and unequilibrated phonons ($\tau\to\infty$, nonequilibrium distribution entirely determined through tunneling dynamics).

It is convenient to reexpress Eq.~\eqref{rateeqs1} in terms of a matrix equation of the form $0=\mathsf{W}\vc{P}$, where $\mathsf{W}$ is a coefficient matrix containing all rates, and the vector $\vc{P}$ consists of all steady-state probabilities. [See Appendix \ref{app:noise} for further details.] For a systematic expansion in orders of the tunneling, we write $\mathsf{W}=\mathsf{W}^{(1)}+\mathsf{W}^{(2)}+\cdots$ as well as $\vc{P}=\vc{P}^{(0)}+\vc{P}^{(1)}+\cdots$. Substituting this into the rate equation, we obtain up to second order the equations
\begin{align}
\mathsf{W}^{(1)}\vc{P}^{(0)}&=0,\\
\mathsf{W}^{(1)}\vc{P}^{(1)}&=-\mathsf{W}^{(2)}\vc{P}^{(0)}.\label{coteqs}
\end{align}
The normalization  condition for $\vc{P}$ leads to $\tr \vc{P}^{(0)}=\sum_i P_i^{(0)}=1$ and $\tr\vc{P}^{(1)}=0$. Here and in the following, the ``trace" of a vector always denotes the sum of its components. Similarly, we expand the steady-state current 
\begin{align}
I&=I^{(1)}+I^{(2)}+\cdots\\\nonumber
&= \tr \mathsf{W}_I^{(1)}\vc{P}^{(0)} + \tr \left[ \mathsf{W}_I^{(1)}\vc{P}^{(1)} + \mathsf{W}_I^{(2)}\vc{P}^{(0)}\right] + \cdots
\end{align}
Here, $\mathsf{W}_I$ denotes the reduced coefficient matrix containing only current-carrying processes, see Appendix \ref{app:noise}. The lowest-order current $I^{(1)}$ is the sequential-tunneling current. The second-order contribution $I^{(2)}$ contains the cotunneling current and corrections to the sequential current due to changes of the probability distribution by cotunneling.

A method for evaluating the current shot noise
\be
S(\omega)=2 \int_{-\infty}^\infty d\tau\, e^{i\omega\tau}\left[ \langle I(\tau)I(0)\rangle - \langle I \rangle^2 \right]
\ee
 within the rate-equations formalism has been developed by Korotkov in Ref.~\onlinecite{korotkov}. This method can be generalized to include higher-order processes beyond sequential tunneling, as we show in detail in Appendix \ref{app:noise}. It is an important question whether the inclusion of negative ``cotunneling rates"  into this noise formalism is justified. In Appendix \ref{app:exact}, we explicitly verify for the resonant-tunneling model that this is indeed the case. The exact solution for the resonant-tunneling model allows for an evaluation of the zero-frequency noise via the relation
\begin{align}
S=\frac{2e^2}{h}\int dE\, &\bigg\{ \sum_a T(E) f_a(E)\left[1-f_a(E)\right]\\\nonumber
&+T(E)\left[ 1-T(E)\right] \left[ f_L(E) - f_R(E) \right]^2
\bigg\},
\end{align}
see Refs.~\onlinecite{lesovik1,blanter}. We find that the leading and next-to-leading order expansion in $\Gamma/k_BT$ of the resulting noise  is exactly reproduced by the generalized Korotkov formalism including negative cotunneling corrections.

\section{Current-voltage characteristics\label{sec:current}}
\subsection{Linear Conductance}
For discrete electronic levels weakly coupled to two electrodes via tunneling junctions, the linear conductance is known to develop peaks whenever an electronic level is aligned with the Fermi energies of the leads. These peaks are well described by sequential-tunneling processes within the rate-equations formalism. Away from the peaks, the sequential-tunneling conductance is exponentially suppressed in the parameter $\epsilon_d/k_BT$, and cotunneling contributions become important. In the following, we show that strong electron-phonon coupling destroys this gate-controlled crossover between sequential tunneling and cotunneling at low temperatures $k_BT\ll\hbar\omega$, rendering cotunneling dominant over wide parameter ranges.

\subsubsection{Sequential-tunneling conductance}
We start by considering the lowest-order contributions to the conductance, in the limit of low temperatures $\Gamma\ll k_BT\ll\hbar\omega$. In this case, only the vibrational ground state is occupied, and any real phonon excitations are negligible. This only leaves the rates $W^{01}_{00}$ and $W^{10}_{00}$ for consideration. Employing the rate-equations approach to this simple situation, the linear conductance for infinite $U$ is obtained as
\be\label{gseq}
G_\text{seq}=-\frac{2e^2}{\hbar}\frac{\Gamma_L\Gamma_R}{\Gamma_L+\Gamma_R}\frac{f'(\epsilon_d)}{1+f(\epsilon_d)}e^{-\lambda^2}.
\ee
In the absence of vibrations, this result can be found, e.g., in Ref.~\onlinecite{pustilnik2}. The crucial point here is the fact that the presence of vibrations leads to an exponential suppression of the linear conductance in $\lambda^2$, even at the peak value. As discussed in Ref.~\onlinecite{pustilnik2}, it is interesting to note that in distinction to the $U=0$ case, the conductance peak position is not at $\epsilon_d=0$ but varies with temperature as $\epsilon_d=k_BT\ln 2/2$. This shift originates from the fact that the sequential-tunneling rates $n=0\to1$ are twice as large as the rates $n=1\to0$ due to the spin degeneracy.

\subsubsection{Cotunneling conductance}
We now consider the next-to-leading order contributions to the conductance. Remaining in the limit of low temperatures $\Gamma\ll k_BT\ll\hbar\omega$, the only two additional processes are elastic cotunneling through the empty and the singly-occupied molecule with corresponding rates $W^{00}_{00;aa'}$ and $W^{11}_{00;aa'}$, respectively. The resulting conductance due to elastic cotunneling is given by
\begin{align}\label{gcot}
G_\text{cot}&=e\frac{\partial}{\partial V}\sum_{n=0,1} P^n_0\left(W^{nn}_{00;LR}-W^{nn}_{00;RL}\right)\bigg|_{V=0}\\\nonumber
&\approx
\frac{2e^2}{h} \frac{\Gamma_L\Gamma_R}{1+f(\epsilon_d)}\bigg[\frac{1-f(\epsilon_d)}{(\epsilon_d+\lambda^2\hbar\omega)^2}
+\frac{2f(\epsilon_d)}{(\epsilon_d-\lambda^2\hbar\omega)^2}\bigg].
\end{align}
Here, the apparent poles at $\epsilon_d=\pm\lambda^2\hbar\omega$ are exponentially suppressed due to the Fermi function factors. In fact, they are an irrelevant artefact of our simple approximation in Eq.~\eqref{cotappr}. The correct treatment via the regularization procedure eliminates this flaw. 

\begin{figure}
	\centering
		\includegraphics[width=0.98\columnwidth]{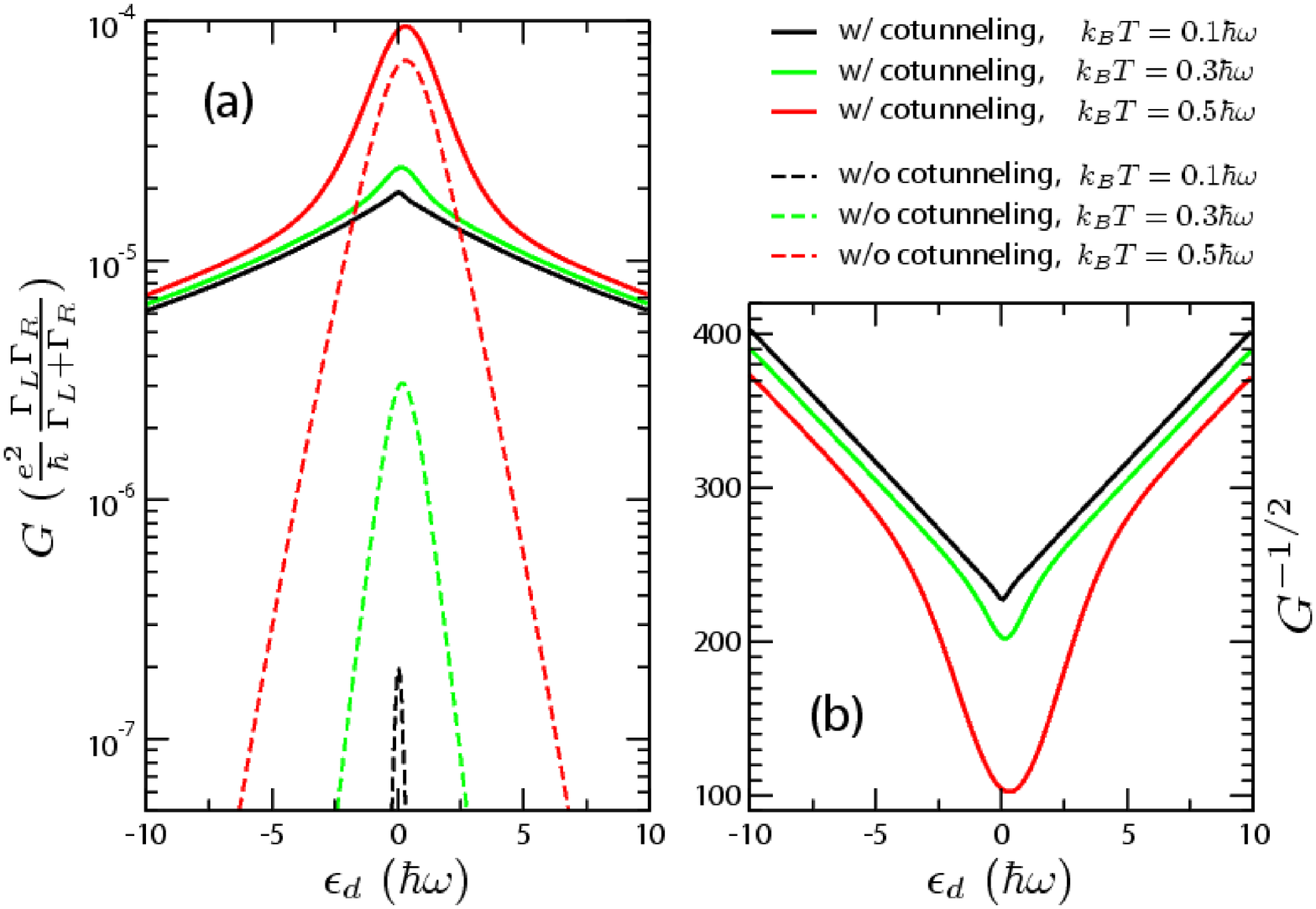}
			\caption{(Color online) Linear conductance as a function of gate voltage in the FC blockade regime ($\lambda=4$, $\Gamma_a=0.02\hbar\omega$). (a) A comparison of the sequential-tunneling conductance (dashed lines) and the conductance including cotunneling corrections (solid lines) clearly shows the dominance of cotunneling contributions at low temperatures. At $\epsilon_d\approx0$, cotunneling results in a conductance kink for $k_BT\ll \hbar\omega$. Sequential contributions become relevant for temperatures $k_BT\agt0.2\hbar\omega$ that allow thermal-induced occupation of excited phonon states. (b) The plot of $G^{-1/2}$ reveals the inverse quadratic decrease of the cotunneling conductance as a function of gate voltage. 
			\label{fig-lincond}}
\end{figure}

By contrast to the $\sim e^{-\lambda^2}$ suppression of the sequential-tunneling conductance, the cotunneling conductance 
is only algebraically suppressed as $\sim\lambda^{-4}$. This is a strong indication that, in the regime of strong electron-phonon coupling, cotunneling will dominate the linear conductance, \emph{even at the conductance peak}. Representative numerical results confirming this expectation are shown in Fig.~\ref{fig-lincond}(a). For $k_BT \ll \hbar\omega$, the cotunneling conductance $G_\text{cot}$ is dominant. In this interesting case, the conductance develops a kink feature at $\epsilon_d=0$  broadened over a gate-voltage range given by $k_BT$, which can be understood from Eq.~\eqref{gcot}. Away from the peak maximum,  $G_\text{cot}$ roughly decays as $\sim(\epsilon_d + \text{const.})^{-2}$, see Fig.~\ref{fig-lincond}(b). This is consistent with Eq.~\eqref{gcot}. For larger temperatures $k_BT \agt 0.2\hbar\omega$, real excitation of phonons becomes important. Such excitations, which we have neglected in our approximations \eqref{gseq} and \eqref{gcot}, open up additional channels for sequential tunneling. The increase of sequential contributions is evident from the high-temperature curve in Fig.~\ref{fig-lincond} (in red/dark grey color), which shows the recurrence of a sequential-tunneling peak on top of the cotunneling background.

\subsection{Nonlinear current-voltage characteristics}
\subsubsection{Sequential current}
For strong electron-phonon coupling, we have established in Ref.\ \onlinecite{koch2} that the nonlinear $IV$ develops a low-bias gap, essentially eliminating the conventional crossing at the degeneracy point, cf.~Fig.~\ref{fig2}(b).
For equilibrated phonons, it is not difficult to estimate the sequential-tunneling current in the FC blockade regime. The main input  is that (i) for low temperatures $k_BT\ll\hbar\omega$ and strong relaxation, the initial vibrational state  is the phonon ground state $q=0$ for each tunneling processes, and  (ii)  the squared Franck-Condon matrix elements $\abs{M_{0q}}^2$ strongly increase with $q$ up to values $q\sim\lambda^2$. 

Due to (ii), the current is dominated by tunneling events with maximal change in the phonon number, $\Delta q=\lfloor \abs{eV}/2\hbar\omega\rfloor$, as long as $\abs{eV}/2\hbar\omega\ll\lambda^2$. Hence, (i) allows us to approximate the current via the rates $W^{01}_{0,\Delta q;\,a}$ as
\be
I_\text{seq}\approx \sgn(V)\frac{e}{\hbar}\frac{2\Gamma_a\Gamma_{a'}}{2\Gamma_a+\Gamma_{a'}}e^{-\lambda^2}\frac{\lambda^{2\Delta q}}{\Delta q!}
\ee
with $(a,a')=(L,R)$ or $(R,L)$ depending on the bias sign. Here, we have assumed $\epsilon_d=0$ for simplicity, which can in principle always be arranged by adjusting the gate voltage. After averaging over steps, one obtains for voltages larger than temperature
\be\label{asymptote}
I(V)\approx\frac{e}{\hbar}\frac{2\Gamma_a\Gamma_{a'}}{2\Gamma_a+\Gamma_{a'}}\frac{e^{-\lambda^2}\lambda^{\abs{eV}/\hbar\omega-1}}{\Gamma(\abs{eV}/2\hbar\omega+1/2)}\sgn(V),
\ee
where the denominator involves the Gamma function $\Gamma(x)$. 

\begin{figure}
	\centering
		\includegraphics[width=0.98\columnwidth]{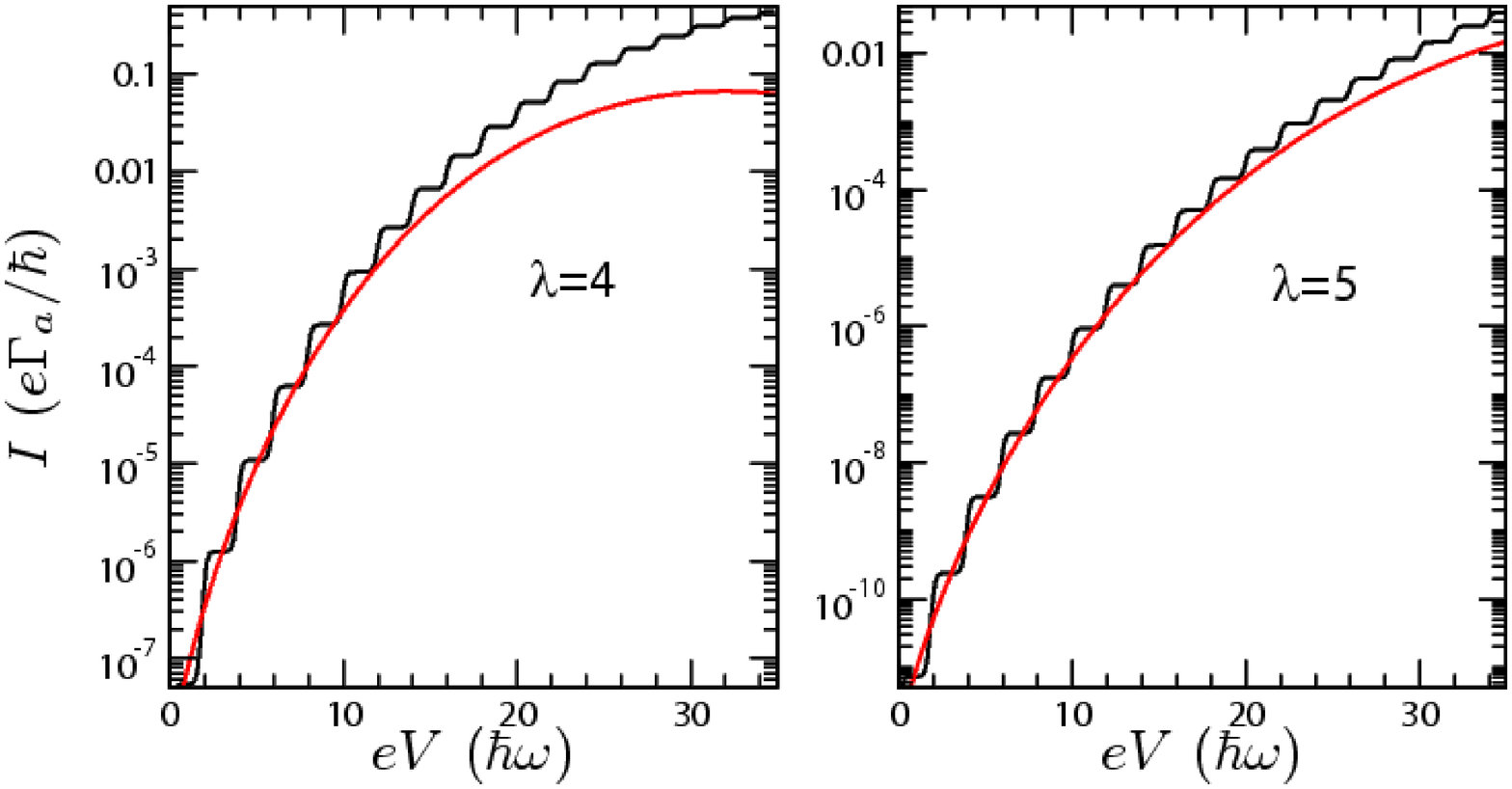}
			\caption{(Color online) Comparison between the sequential-tunneling current [black curve] and the asymptotic formula  \eqref{asymptote} [red/grey curve] for equilibrated phonons in the Franck-Condon blockade regime. ($\epsilon_d=0$, $k_BT=0.05\hbar\omega$, $\Gamma_L=\Gamma_R$)
			\label{fig-eqasympt}}
\end{figure}

The comparison between the asymptotic form Eq.~\eqref{asymptote} and the actual rate-equations solution is depicted in Fig.~\ref{fig-eqasympt}. Deviations of the asymptotic form from the actual solution are observed when the inequality $2\abs{eV} \ll \lambda^2\hbar\omega$ gradually breaks down. Then, the true current assumes higher values as compared to Eq.~\eqref{asymptote}, since assumption (ii) is no longer valid, i.e.\ not only one but several phonon channels yield significant contributions to the current.

For unequilibrated phonons, where FC blockade leads to self-similar avalanches of electrons,\cite{koch2,koch3} an estimate for the sequential current is less trivial. The avalanche dynamics makes it necessary to include several phonon channels. In particular, following the arguments of Ref.~\onlinecite{koch3}, the mean electron-number per avalanche as well as the mean number of sub-avalanches within each avalanche will depend on bias and gate voltage, as well as on the coupling strength $\lambda$.

\subsubsection{Nonlinear $IV$ including cotunneling corrections}
\begin{figure}
	\centering
		\includegraphics[width=0.9\columnwidth]{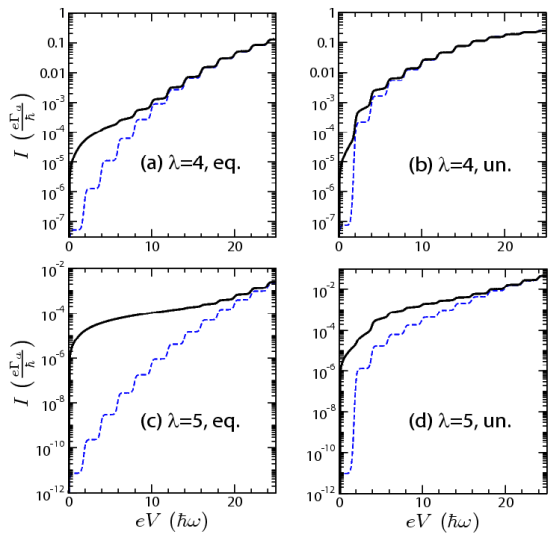}
	\caption{(Color online)  Nonlinear current-voltage characteristics in the FC blockade regime. Each graph shows the steady-state current $I^{(1)}+I^{(2)}$ (solid curves), and the sequential current $I^{(1)}$ (dashed curves) for comparison. The electron-phonon coupling and vibrational relaxation strength are chosen as (a) $\lambda=4$, equilibrated, (b) $\lambda=4$, unequilibrated, (c) $\lambda=5$, equilibrated, (d) $\lambda=5$, unequilibrated. The results unambiguously indicate the dominance of cotunneling contributions in the low-bias FC suppressed regime. (Parameters: $\epsilon_d=0$, $k_BT=0.05\hbar\omega$, $\Gamma_a=0.02\hbar\omega$)\label{fig1}}
\end{figure}

\begin{figure}
	\centering
		\includegraphics[width=0.99\columnwidth]{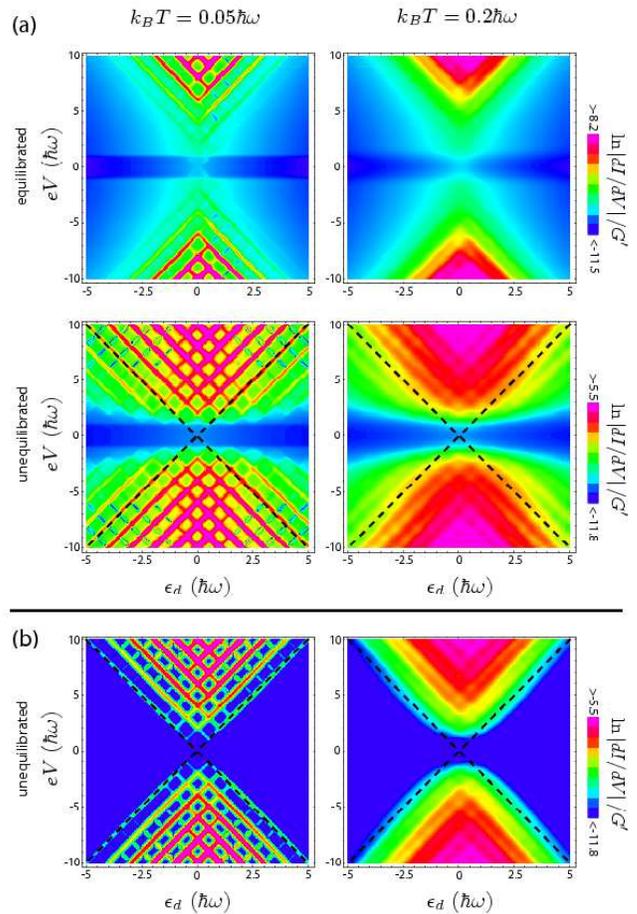}
	\caption{(Color online)  (a) Differential conductance $dI/dV$ as a function of gate and bias voltage for $\lambda=4$ and $\Gamma_a=0.02\hbar\omega$, plotted with a logarithmic color scale. The low temperature plots (left-hand side) show a step in $dI/dV$ due to the onset of cotunneling (horizontal feature at $eV=\pm\hbar\omega$). For unequilibrated phonons, inelastic cotunneling gives rise to phonon absorption-induced vibrational sidebands inside the Coulomb-blockade region (whose borders are marked by dashed lines). For comparison, the plots in (b) show the corresponding results within the sequential-tunneling approximation which misses the absorption-induced sidebands. ($G'=\frac{e^2}{\hbar}\frac{\Gamma_L\Gamma_R}{\Gamma_L+\Gamma_R}$)\label{fig2}}
\end{figure}

We now investigate the cotunneling corrections to the current-voltage characteristics in the FC blockade regime. Representative results for both equilibrated and unequilibrated vibrations are depicted in Figs.~\ref{fig1} and \ref{fig2}.

We first focus on the case of \emph{equilibrated phonons}. The $IV$ plots for $\epsilon_d=0$, Fig.~\ref{fig1}(a) and (c), illustrate the importance of cotunneling contributions well beyond the linear response regime. Cotunneling corrections significantly increase the current for bias voltages $\abs{eV}< \lambda^2\hbar\omega$, the reason being again the increased overlap for oscillator states in cotunneling processes with highly excited intermediate states. It is a crucial point, however, that cotunneling does not lead to a breakdown of the FC blockade. The blockade remains intact, since cotunneling rates are still algebraically suppressed at strong electron-phonon coupling. This suppression remains strong due to the rather large exponents in the scaling with $\lambda$, see Eq.~\eqref{cotappr}. 
An additional change caused by the dominance of cotunneling rates is a drastic reduction of step features in the current, see especially Fig.~\ref{fig1}(c). When plotting the differential conductance $dI/dV$ on a logarithmic scale, see Fig.~\ref{fig2}, one observes that small steps persist on top of the (much larger) cotunneling background and form the usual system of vibrational sidebands.

An important low-bias feature beyond the sequential-tunneling approximation consists of two steps in $dI/dV$ for $eV=\pm\hbar\omega$. These steps mark the onset of inelastic cotunneling. The emergence of such steps is not restricted to the FC blockade, and has been discussed, e.g., in Ref.~\onlinecite{varma}. Additional steps at bias voltages equal to larger integer multiples of $\hbar\omega$ can be observed when going to even lower temperatures than chosen in Figs.~\ref{fig1} and \ref{fig2}.

For \emph{unequilibrated phonons}, we find that cotunneling corrections again dominate the FC blockade region of the current-voltage characteristics, see Fig.~\ref{fig1}(b),(d). However, relative to the equilibrated case, the corrections are less drastic. In particular, vibrational steps in $IV$ remain well visible, and the bias range where cotunneling dominates is  narrower than in the equilibrated case. This can be explained by the persistence of avalanche dynamics generated by sequential tunneling. Within the FC blockade, important contributions to the current stem from rather featureless cotunneling events. However, phases of pure cotunneling are still interrupted by avalanches.  During those, the system attains phonon excitations high enough to make sequential tunneling dominate over cotunneling. Hence, sequential tunneling retains a significant role in the unequilibrated case.

Fig.\ \ref{fig2} demonstrates an additional striking consequence of cotunneling: For unequilibrated phonons, additional vibrational sidebands appear in the Coulomb-blockade region. The mechanism which generates these additional sidebands is the following.
The Coulomb-blockade regime, where transport is conventionally dominated by cotunneling, allows for excitation of vibrations through inelastic cotunneling as soon as the bias voltage exceeds the phonon energy. For weak vibrational relaxation, such excitations decay slowly and therefore remain relevant for subsequent tunneling events. The crucial point is that the usual exponential suppression for sequential-tunneling in the Coulomb blockade applies to processes which leave the vibrational state \emph{unchanged},\footnote{More generally, the suppression applies to all sequential processes which leave the vibrational state unchanged or increase the excitation level.} but not to processes gaining energy through phonon deexcitation. Thus, phonon excitations (induced by inelastic cotunneling) facilitate the recurrence of sequential tunneling accompanied by phonon deexcitation in the Coulomb-blockaded region. Furthermore, the onset of inelastic cotunneling at $\abs{eV}=\hbar\omega$ and the resulting absence of vibrational excitations for lower bias voltages explain why absorption-induced sidebands must terminate at $\abs{eV}=\hbar\omega$ and cannot enter the low-bias region.

This interplay between inelastic cotunneling and sequential tunneling results in the observed absorption-induced vibrational sidebands,\footnote{The behavior of these absorption-induced vibrational sidebands, which are not specific to the case of strong electron-phonon coupling, deserves further investigations, and will be published elsewhere. (M.\ C.\ L\"uffe, J.\ Koch, and F.\ von Oppen, in preparation)} resembling experimental findings in transport through suspended carbon-nanotubes.\cite{LeRoy} We emphasize that the absorption-induced vibrational sidebands are \emph{not} specific to the case of strong electron-phonon coupling, and point out that their presence may be exploited to estimate the vibrational relaxation rate in molecular junctions.\cite{LeRoy}

Interestingly, phonon excitation due to inelastic cotunneling is the only viable mechanism for absorption-induced vibrational sidebands up to second order in the tunneling. It is crucial to note that thermally induced phonon excitations will in general \emph{not} lead to visible additional sidebands, but rather to a smooth smearing of the Coulomb diamond, see Fig.~\ref{fig2}(b). This is explained by the fact that vibrational sidebands are clearly visible only for $k_BT\ll\hbar\omega$, and in this case thermally activated phonon excitations are negligible.

\section{Current shot noise\label{sec:noise}}
The current shot noise serves as an important tool for acquiring information about the transport dynamics of the system beyond the steady-state current. In particular, we are interested in the Fano factor $F$ and noise spectrum $S(\omega)$ for unequilibrated vibrations to determine the fate of the self-similar avalanches in the presence of cotunneling corrections. We first turn to the regime of strong  vibrational relaxation and show that even in this case, cotunneling leads to interesting new effects. 

\subsection{Zero-frequency noise for equilibrated vibrations: Telegraph noise}
\begin{figure}
	\centering
    \vspace*{0.5cm}
		\includegraphics[width=0.98\columnwidth]{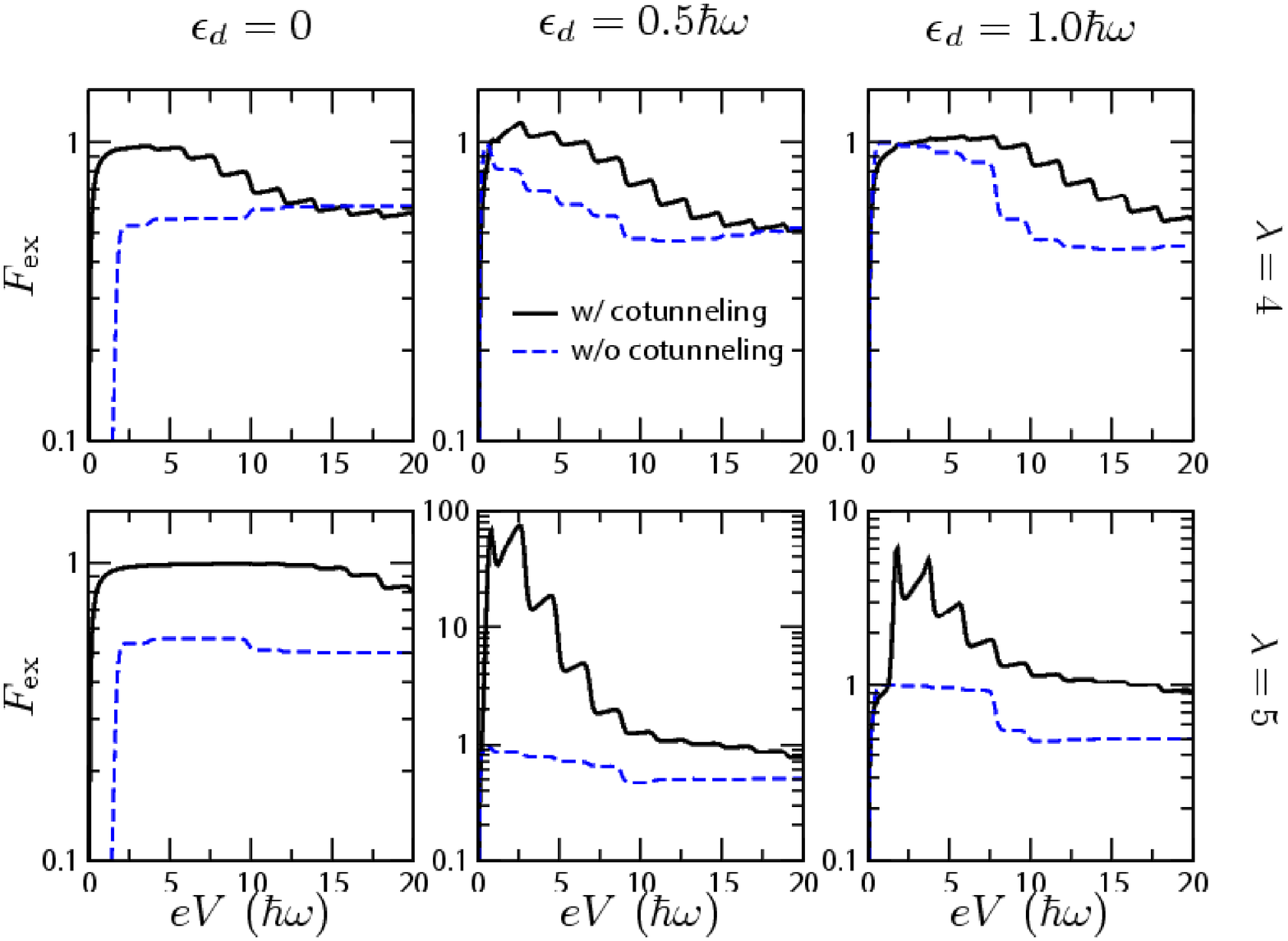}
			\caption{(Color online) Excess noise Fano factor for equilibrated phonons with $\lambda=4$ (top row) and $\lambda=5$ (bottom row). For nonzero $\epsilon_d$ we observe large super-Poissonian Fano factors due to current telegraph noise. ($k_BT=0.05\hbar\omega$, $\Gamma_a=0.02\hbar\omega$)
			\label{fig-fanoeq}}
\end{figure}

As shown in Ref.~\onlinecite{koch2},  sequential tunneling results in conventional sub-Poissonian Fano factors for equilibrated phonons, see the dashed curves in Fig.~\ref{fig-fanoeq}. Electrons are transferred one by one,  interacting with each other only via the charging energy $U$, which leads to the suppression of $F$ below $1$. As depicted in Fig.~\ref{fig-fanoeq} by the black curves, the inclusion of cotunneling processes radically changes this picture. The overall Fano factor remains sub-Poissonian for $\epsilon_d=0$. Remarkably, it switches to super-Poissonian behavior for nonzero $\epsilon_d$ in a certain bias range. The magnitude of $F$ strongly depends on the coupling strength $\lambda$, as well as gate and bias voltage, and reaches values as large as $100$ for $\lambda=5$.

More insight into this surprising enhancement of noise, which even occurs in the low-bias regime where no phonons can be excited, is obtained by Monte-Carlo simulations. For the parameters of the large maximum of the Fano factor in Fig.~\ref{fig-fanoeq} ($\lambda=5$, $\epsilon_d=0.5\hbar\omega$), Fig.\ \ref{fig-telegraph} shows the coarse-grained current as a function of time.  These data clearly show the slow switching between a low and a high current state, on top of much more rapid dynamics within each current state. The underlying reason for this telegraph noise is the role inversion of sequential-tunneling and cotunneling rates in the FC blockade, i.e.\ rates for cotunneling are larger than those for sequential tunneling.

In the following, we give a detailed explanation of the telegraph behavior. Denoting the molecular state by $\ket{n,q}$, a straightforward inspection shows that for $\epsilon_d=0.5\hbar\omega$ and $eV=0.8\hbar\omega$, the only relevant transitions are (1) $\ket{0,0}\to\ket{1,0}$ and $\ket{1,0}\to\ket{0,0}$ due to sequential tunneling, and (2) $\ket{0,0}\to\ket{0,0}$ and $\ket{1,1}\to\ket{1,1}$ due to cotunneling. As explained in Section \ref{sec:formalism}, the cotunneling transitions are much faster than the sequential transitions in the FC blockade regime, such that $W^{00}_{00},\,W^{11}_{00} \gg W^{01}_{00},\,W^{10}_{00}$. The key point in the emergence of the telegraph behavior is the fact that the rates for cotunneling through the empty and through the singly occupied dot are generally different, $W^{00}_{00}\not=W^{11}_{00}$. This can be understood from Eqs.~\eqref{cotrate1} and \eqref{cotrate2}, noting that $\epsilon_d$ enters the energy denominator with a positive sign for $n=1$, but with a negative sign for $n=0$. For $\epsilon_d=0$, the two cotunneling rates are identical, and indeed no super-Poissonian noise is detected in this case. However, for nonzero $\epsilon_d$, the two rates become different. As a consequence, the rapid cotunneling dynamics in the charge states $n=0,1$ leads to different mean currents $eW^{00}_{00}$ and $eW^{11}_{00}$, respectively. The rare sequential transitions effect the slow switching between these two current states.

The Fano factor resulting from the current telegraph noise is directly related to the excess charge transmitted in the high-current state during the typical residence time in the corresponding charge state, i.e.\ $F\sim\Delta W_\text{cot}/W_\text{seq}$. Here, $\Delta W_\text{cot}=\abs{W^{00}_{00}-W^{11}_{00}}$ denotes the difference of the relevant cotunneling rates. Due to the role inversion of cotunneling and sequential tunneling, we have $W_\text{cot}\gg W_\text{seq}$ so that the resulting Fano factor may become large.
A more quantitative calculation of the Fano factor can be achieved by employing the generalization of Korotkov's method\cite{korotkov} described in Appendix \ref{app:noise}. Along these lines,  we can derive approximate expressions for the zero-frequency shot noise and its Fano factor,
\begin{align}
S(f=0)&\approx 4e^2\frac{W^{01}_{00}W^{10}_{00}(W^{00}_{00}-W^{11}_{00})^2}{(W^{01}_{00}+W^{10}_{00})^3},\\
F&\approx 6\frac{W^{01}_{00}W^{10}_{00}(W^{00}_{00}-W^{11}_{00})^2}{(W^{00}_{00}+2W^{11}_{00})(W^{01}_{00}+W^{10}_{00})^3},
\end{align}
see Appendix \ref{app:telegraph} for details. This approximation is valid whenever $W^{00}_{00},\,W^{11}_{00} \gg W^{01}_{00},\,W^{10}_{00}$ and $W^{00}_{00}\not=W^{11}_{00}$. Counting the orders of cotunneling versus sequential-tunneling rates in the numerator and denominator of the Fano factor, we find that the Fano factor is indeed of the order of $\Delta W_\text{cot}/W_\text{seq}$.

\subsection{Zero-frequency noise for unequilibrated vibrations: Persistence of avalanches}
\begin{figure}
	\centering
    \vspace*{0.5cm}
		\includegraphics[width=0.98\columnwidth]{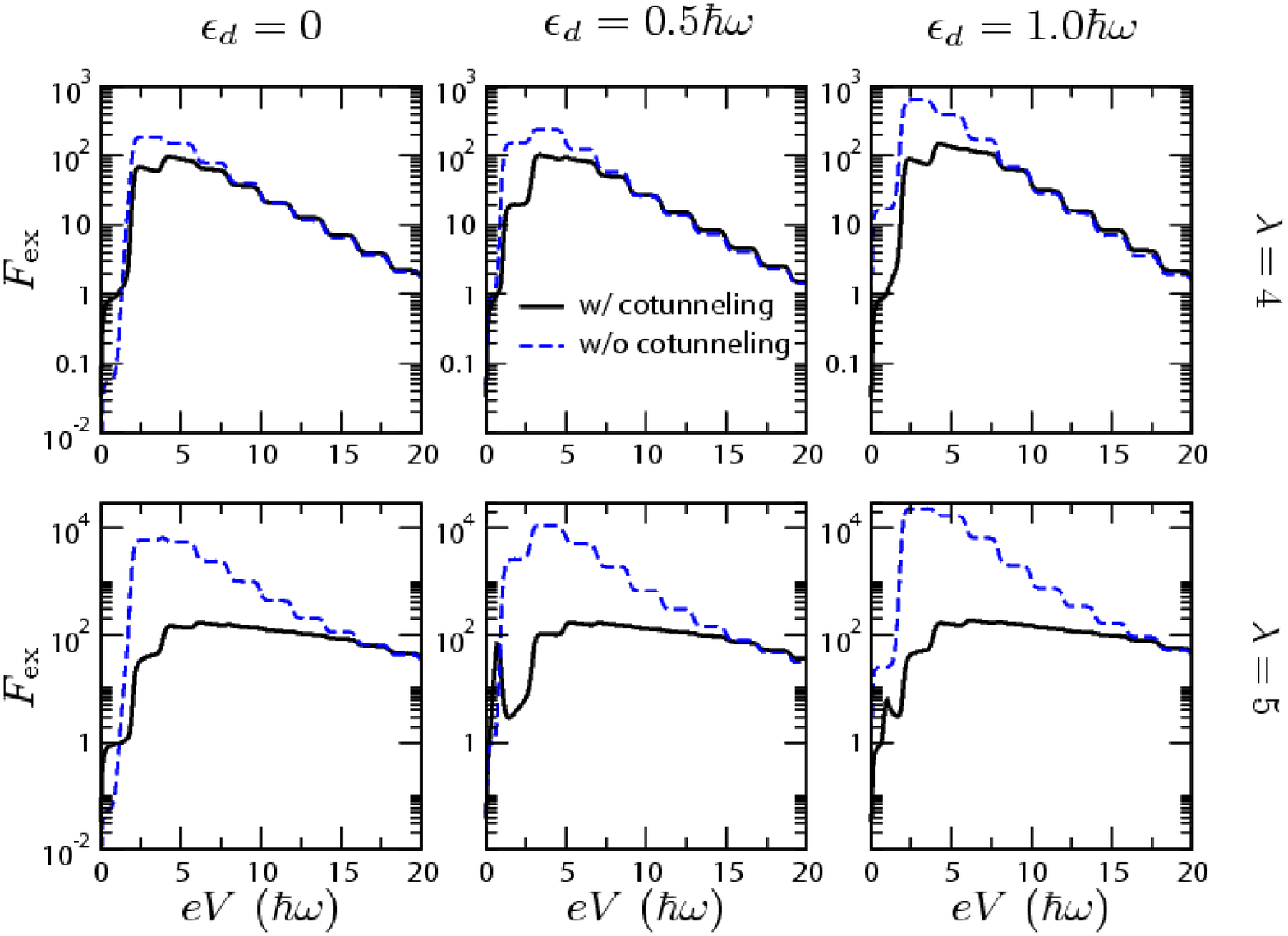}
			\caption{(Color online) Excess noise Fano factor for unequilibrated phonons with $\lambda=4$ (top row) and $\lambda=5$ (bottom row) with $k_BT=0.05\hbar\omega$ and $\Gamma_a=0.02\hbar\omega$. Due to the coexistence of cotunneling and avalanches, the Fano factor is reduced as compared to the sequential-tunneling approximation. The super-Poissonian peak observed at low bias for $\lambda=5$ and $\epsilon_d\not=0$ signals the occurrence of current telegraph noise.
			\label{fig-fanoun}}
\end{figure}

For unequilibrated vibrations, the sequential-tunneling dynamics in the FC blockade regime is described by self-similar avalanches of electrons.\cite{koch2,koch3} The large Fano factors due to sequential tunneling shown in Fig.~\ref{fig-fanoun} (dashed curves) reflect the large number of electrons per avalanche. In the $\epsilon_d=0$ case, the Fano factor is exactly given by $F=2\langle N_i \rangle$, i.e.\ twice the mean number of electrons per level-0 avalanche. More generally, the Fano factor can be obtained by evaluating the fluctuations of the waiting times $t_i$ between level-0 avalanches, and the fluctuations of the electron number $N_i$ in each avalanche,\cite{koch3}
\be\label{fanogen}
  F=\langle N_i\rangle \frac{\langle t_i^2\rangle-\langle t_i\rangle^2}{\langle t_i\rangle^2}
          +\frac{\langle N_i^2\rangle-\langle N_i\rangle^2}{ \langle N_i\rangle}.
\ee%

Before turning to the cotunneling-induced corrections, we study the effect of nonzero $\epsilon_d$, which was not discussed in Refs. \onlinecite{koch2} and \onlinecite{koch3} in detail. Interestingly, a nonzero $\epsilon_d$ can lead to even larger Fano factors as compared to the $\epsilon_d=0$ case, see Fig.~\ref{fig-fanoun}. Especially large Fano factors are observed for low biases above the threshold $eV>\hbar\omega$ whenever the electronic level $\epsilon_d$ is roughly aligned with the Fermi energy of the right or left lead.

At first view, the resulting Fano factor enhancement is surprising since the mean number of electrons per avalanche is \emph{decreased} for such a level configuration. For an understanding of this decrease, consider the case of level alignment with the left Fermi energy. Then, phonon excitations can only occur through tunneling in the right junction. Events in the junction $L$ can only decrease the phonon state (or leave it unchanged). Thus, every second tunneling event will inhibit phonon excitations. With avalanches benefitting from the possibility of reaching highly excited phonon states, the mean number of electrons per avalanche is reduced as compared to the case where the level is centered between the two Fermi levels. This decrease in $\langle N_i\rangle$ is confirmed by Monte-Carlo simulations. 

The somewhat counterintuitive result that the Fano factor nevertheless becomes \emph{larger} can be traced back to the significantly enhanced fluctuations in waiting times $t_i$, which directly contribute to the Fano factor, see Eq.~\eqref{fanogen}. In the following, we explain this enhancement of waiting-time fluctuations.
Any generation--0 avalanche may terminate either in the charge state $n=0$ or $n=1$, and the corresponding probabilities are denoted $p_0$ and $p_1$. After the avalanche termination, the system undergoes a waiting period $t_i$. The crucial point is that for level alignment with one Fermi energy, this waiting time strongly depends on the charge state $n$. Starting in the state $n=0$, transitions in the left junction do not allow phonon excitations. By contrast, for the initial state $n=1$ and transitions across the right junction, the excitation of phonons is favored (due to the increase of FC matrix elements towards highly excited vibrational states). 
Accordingly, the relevant two rates after avalanche termination are dramatically different,  $W^{01}_{00}\ll W^{10}_{01}$, and the waiting times in the two cases will differ accordingly,
\be
\langle t_i^{(n=0)}\rangle = (W^{01}_{00})^{-1} \gg  \langle t_i^{(n=1)}\rangle = (W^{10}_{01})^{-1}.
\ee
Consequently, the probability distribution of waiting times is not given by a single exponential anymore, but is well approximated by
\be
p(t_i)\approx p_0 W^{01}_{00} \exp[-W^{01}_{00}t_i]+ p_1 W^{10}_{01} \exp[-W^{10}_{01}t_i].
\ee
While the $\epsilon_d=0$ case (with a simple exponential distribution) results in $\left( \langle t_i^2\rangle-\langle t_i\rangle^2\right)/\langle t_i\rangle^2=1$, we now find
\be\label{tfluc}
 \frac{\langle t_i^2\rangle-\langle t_i\rangle^2}{\langle t_i\rangle^2}\approx \frac{2-p_0}{p_0},
\ee
valid for $p_0\gg W^{01}_{00}/W^{10}_{01}$. By Monte-Carlo simulations we confirm that $p_0$ satisfies $1>p_0\gg W^{01}_{00}/W^{10}_{01}$ for realistic parameters. Therefore, Eq.~\eqref{tfluc} explains the enhancement of Fano factor due to the increased waiting-time fluctuations.

The inclusion of cotunneling contributions to the current shot noise results in the black curves depicted in Fig.~\ref{fig-fanoun}. Again, at low biases which do not allow real phonon excitations, we observe traces of telegraph noise. [See the $\lambda=5$ plots for nonzero $\epsilon_d$ in Fig.~\ref{fig-fanoun}.] For higher biases which facilitate avalanche dynamics, we find that cotunneling generally reduces the Fano factor as compared to pure sequential tunneling. The reason for this reduction is the coexistence of electron avalanches (leading to large Fano factors) with cotunneling, which results in Fano factors of the order of unity away from the telegraph noise regime. We emphasize that despite the cotunneling-induced noise reduction, the overall Fano factor remains large compared to unity, signalling the persistence of electron bunching in avalanches. 

\subsection{Noise spectrum for unequilibrated vibrations: Self-similarity of avalanches}
\begin{figure}
	\centering
    \vspace*{0.5cm}
		\includegraphics[width=0.95\columnwidth]{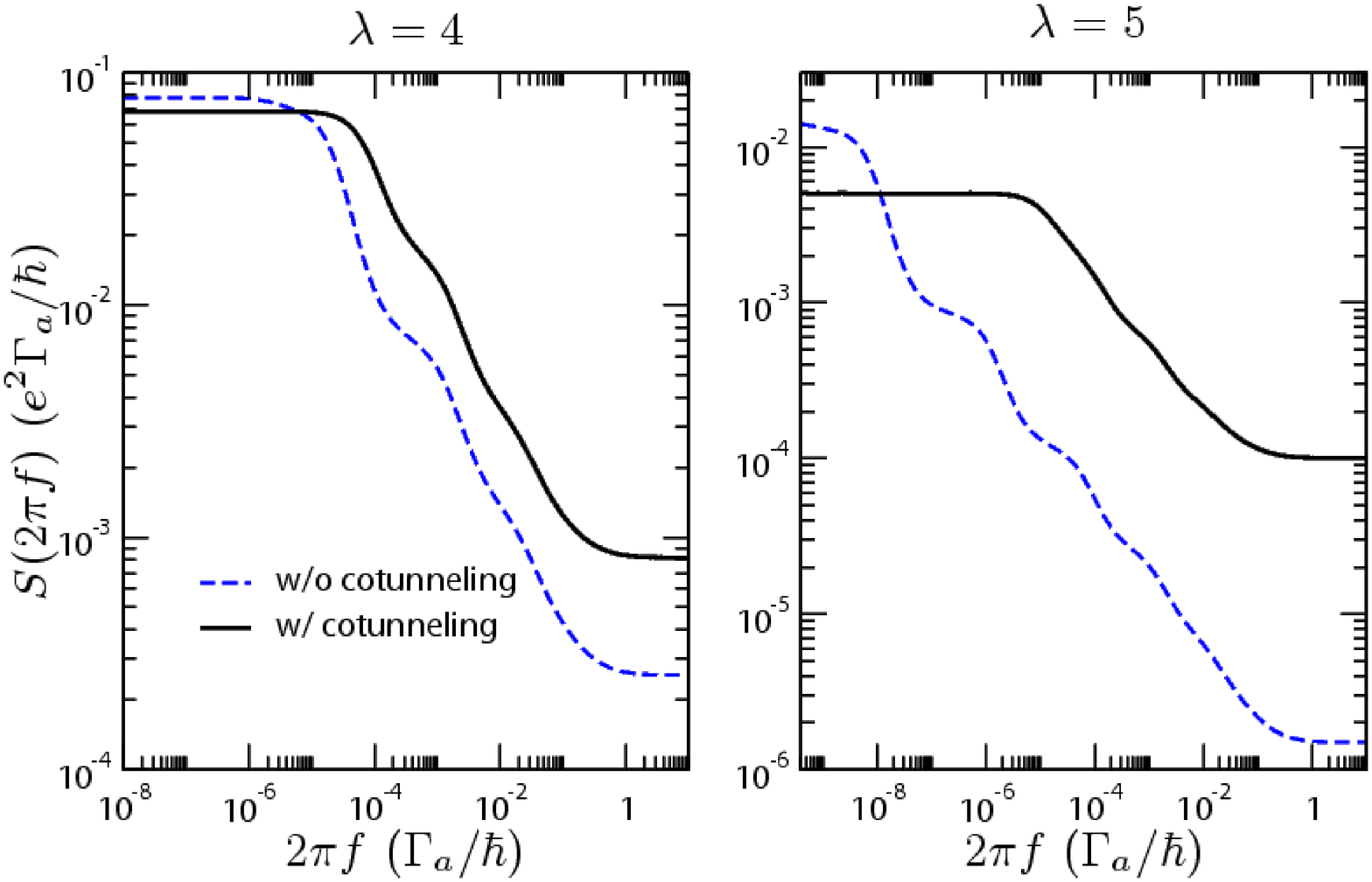}
			\caption{(Color online) Noise power spectrum for unequilibrated phonons with $\lambda=4$ (left) and $\lambda=5$ (right) with $k_BT=0.05\hbar\omega$, $\Gamma_a=0.02\hbar\omega$, $\epsilon_d=0$, and $eV=3\hbar\omega$. The approximate power-law behavior reflecting the self-similarity of avalanches persists in the presence of cotunneling. The dominance of cotunneling during sequential waiting times provides a long-time (low-frequency) cutoff for the power law.
			\label{fig-spectrum}}
\end{figure}

Probing the self-similarity of avalanches requires one to go beyond the zero-frequency noise. In Refs.~\onlinecite{koch2,koch3} two of us have shown that the characteristic signature of the self-similarity within the sequential-tunneling approximation is an approximate power-law behavior of the noise as a function of frequency $f$, $S\sim f^{-1/2}$. In Figure \ref{fig-spectrum} we present the results for the noise power spectrum including cotunneling corrections.

The central result is that the power-law scaling remains valid. Only the frequency range over which it applies is reduced by cotunneling. In particular, we observe a larger low-frequency cutoff when taking cotunneling into account. This result is consistent with the picture of coexistence of cotunneling and sequential avalanches: In the sequential-tunneling picture, the onset of the power law at low frequencies is essentially given by the strongly suppressed rate for escaping from the vibrational ground state, i.e.\ $f\sim W^{01/10}_{01}$. Going beyond sequential tunneling, we have shown that transitions close to the vibrational ground state are dominated by cotunneling which does not contribute to the avalanche dynamics. Accordingly, the onset of self-similarity and hence the power-law behavior is shifted towards higher phonon levels with corresponding higher rates. This is reflected by a larger low-frequency cutoff for the power law in the noise spectrum.

\section{Conclusions\label{sec:conclusions}}
In conclusion, we have developed a complete theory of the Franck-Condon blockade including cotunneling corrections. This regime, caused by a strong coupling to the bosonic degrees of freedom of molecular vibrations, emphasizes the stark contrast between transport through molecular junctions and more conventional nanostructures such as quantum dots. In particular, the intriguing finding that electron-phonon coupling may result in a significant bunching of electrons into a hierarchy of self-similar avalanches dominating the transport, constitutes a novel collective effect unknown from electron transfer through quantum dots. 

In distinction to the Coulomb blockade scenario in quantum dots, where the crossover between sequential tunneling at the conductance peaks and cotunneling away from the peaks may be probed by tuning the gate voltage, we find that the FC blockade regime features the dominance of cotunneling contributions in the entire low-bias region. The underlying reason for this result is the enhanced overlap between vibrational state, which is exploited by highly excited vibrational states in cotunneling processes. We have shown that the FC blockade and avalanche-type transport persist despite the cotunneling corrections. Furthermore, we have established that cotunneling leads to a number of remarkable effects missed by the sequential-tunneling approximation. In summary, our central results are: (i) Apart from horizontal $dI/dV$ steps due to inelastic cotunneling, the most important consequence of cotunneling is the appearance of additional absorption-induced vibrational sidebands inside the Coulomb-blockade regime, relevant for unequilibrated vibrations. This may also serve as an experimental measure of the vibrational relaxation rate. (ii) The role inversion between cotunneling and sequential tunneling may cause current telegraph noise whenever the electronic level is not centered between the left and right Fermi energy. This is reflected in strongly enhanced Fano factors, independet of the vibrational relaxation strength. 

For a more detailed summary of results we refer the reader to the summary provided in Section \ref{sec:intro}. In closing, we emphasize the appeal of realizing shot noise measurements in experiments with single-molecule devices. These could provide essential information about the transport mode, which cannot be accessed by current-voltage characteristics. Experimental efforts in this direction are under way in several research groups.

\begin{acknowledgments}
We would like to thank M.\ E.\ Raikh, M.\ R.\ Wegewijs and K.\ A.\ Matveev for valuable discussions. We gratefully acknowledge hospitality by the University of Washington (JK) and the Weizmann Institute of Science (JK and FvO), made possible by the EU - Transnational Access program (RITA-CT-2003-506095). This work was supported in part by the DFG through Sfb 658 and Spp 1243, by the Studienstiftung des dt.~Volkes, by the German-Israeli Project Cooperation (DIP) and by the David and Lucille Packard Foundation.
\end{acknowledgments}

\appendix
\section{Regularized cotunneling rates\label{app:cotrates}}
The regularized cotunneling rates can be expressed in terms of digamma and trigamma functions,\cite{koch}
\begin{widetext}
\allowdisplaybreaks
\begin{align}\label{reg1}
W^{0\to 0}_{q\to q';ab}=\frac{2\Gamma_a\Gamma_b}{2\pi\hbar} &\bigg[ \sum_r \abs{M_{q'r}M_{qr}}^2 J\left(\mu_a,\mu_b-[q-q']\hbar\omega,\epsilon_d-[q-r]\hbar\omega\right)\\\nonumber
&+\sum_{r\not=s} M_{q'r}M_{qr}^*M_{q's}^*M_{qs} I\left(\mu_a,\mu_b-[q-q']\hbar\omega,\epsilon_d-[q-r]\hbar\omega,\epsilon_d-[q-s]\hbar\omega\right)
\bigg],\\
W^{1\to 1}_{q\to q';ab}=\frac{2\Gamma_a\Gamma_b}{2\pi\hbar} &\bigg[ \sum_r \abs{M_{q'r}M_{qr}}^2 J\left(\mu_a,\mu_b-[q-q']\hbar\omega,\epsilon_d+[q'-r]\hbar\omega\right)\\\nonumber
&+\sum_{r\not=s} M_{q'r}M_{qr}^*M_{q's}^*M_{qs} I\left(\mu_a,\mu_b-[q-q']\hbar\omega,\epsilon_d+[q'-r]\hbar\omega,\epsilon_d+[q'-s]\hbar\omega\right)
\bigg].\label{reg2}
\end{align}
Here, the regularized expressions $I$ and $J$ can formally be obtained by evaluating the $\gamma\to0$ limit after subtracting $1/\gamma$ terms,
\begin{align}
I(E_1,E_2,\epsilon_1,\epsilon_2)&=\lim_{\gamma\rightarrow0} \rre\int dE\, f(E-E_1)\left[ 1-f(E-E_2)\right]\frac{1}{E-\epsilon_1-i\gamma}\frac{1}{E-\epsilon_2+i\gamma}\\\nonumber
&=\frac{n_B(E_2-E_1)}{\epsilon_1-\epsilon_2}\rre\bigg\{ \psi(1/2+i\beta[E_2-\epsilon_1]/2\pi) -\psi(1/2-i\beta[E_2-\epsilon_2]/2\pi) \\\nonumber
&\qquad\qquad\qquad\qquad   - \psi(1/2+i\beta[E_1-\epsilon_1]/2\pi) +\psi(1/2-i\beta[E_1-\epsilon_2]/2\pi) \bigg\}\\
J(E_1,E_2,\epsilon)&=\lim_{\gamma\rightarrow 0} \left[  \int dE\, f(E-E_1)\left[1-f(E-E_2)\right]\frac{1}{(E-\epsilon)^2+\gamma^2}-\mathcal{O}(1/\gamma)\right]\\\nonumber
&=\frac{\beta}{2\pi}n_B(E_2-E_1) \iim \left\{  \psi'\left(1/2+i\beta[E_2-\epsilon]/2\pi\right) - \psi'\left(1/2+i\beta[E_1-\epsilon]/2\pi\right)\right\}.
\end{align}
\end{widetext}
Our numerical implementation of the polygamma function $\psi$ has been based on a routine written by K\"olbig, available in the CERN Program Library.\cite{koelbig,cernlib}

\section{Exactly solvable model\label{app:exact}}
We consider the simple case of a single spin-degenerate, noninteracting level coupled to two leads. Due to the absence of interaction in this model,  spin only results in trivial factors of 2 and may be neglected. 
We have picked this trivial model, since it is exactly solvable, e.g.\ using a one-particle scattering approach.  

\subsection{Current}
The current is given by
\be\label{landauer}
I=\frac{e}{h}\int dE\, \frac{\Gamma_L(E)\Gamma_R(E)}{\Gamma_L(E)+\Gamma_R(E)}A_d(E)[f_L(E)-f_{R}(E)],
\ee
wfere $A_d$ denotes the spectral function of the dot,
\be
A_d(E)=-2\iim G^R_d(E)=\frac{\Gamma(E)}{[E-\epsilon_d]^2+[\Gamma(E)/2]^2},
\ee
and the energy broadening of the level is $\Gamma(E)=\Gamma_L(E)+\Gamma_R(E)$ with $\Gamma_a(E)=2\pi\rho_a(E)\abs{t_a}^2$. (Note that any possible energy shift due to the tunnel coupling has been absorbed into $\epsilon_d$.) In the wide-band limit and for symmetric voltage splitting, the solution reads
\begin{align}\nonumber
I&=\frac{e}{h}\frac{\Gamma_L\Gamma_R}{\Gamma_L+\Gamma_R}\int dE \frac{\Gamma}{[E-\epsilon_d]^2+[\Gamma/2]^2}\left[f_L(E)-f_R(E)\right]\\\label{FINT}
&=\frac{e}{\hbar}\frac{\Gamma_L\Gamma_R}{\Gamma_L+\Gamma_R}\frac{1}{\pi}\iim \sum_a a\psi\left( \frac{1}{2}+\frac{\Gamma\beta}{4\pi}+i\frac{\beta}{2\pi}[\epsilon_d+aeV/2]\right).
\end{align}
Here, $\psi$ denotes the digamma function. The exact result can be expanded in orders of $\Gamma$, which gives
\begin{align}\label{IfinTx}
&I=\frac{e}{\hbar}\frac{\Gamma_L\Gamma_R}{\Gamma_L+\Gamma_R}\left[ f_L(\epsilon_d)-f_R(\epsilon_d)\right]\\\nonumber
&+\frac{e}{h}\frac{\beta}{2\pi}\Gamma_L\Gamma_R
\iim \sum_a a \psi'\left( \frac{1}{2}+i\frac{\beta}{2\pi}[\epsilon_d+aeV/2]\right)+\mathcal{O}(\Gamma^3)
\end{align}
For the weak-coupling case we can alternatively apply the rate-equations approach. In this case, the stationary current is calculated via $I=I_L=-I_R$ with
\begin{align}\label{current}
I_L&=I_L^\text{(seq)}+I_L^\text{(cot)}\\\nonumber
&=e[P_0 W^+_L -P_1 W^-_L ] \\\nonumber
&\quad+ e[P_0(W^{00}_{LR} -W^{00}_{RL})+P_1(W^{11}_{LR} -W^{11}_{RL})].
\end{align}
Here, $W^+_a$ ($W^-_a$) gives the rate for a sequential-tunneling process in junction $a$ where one electron enters (leaves) the dot. Similarly, $W^{nn}_{LR}$ ($W^{nn}_{RL}$) denotes the rate for a cotunneling process with initial and final charge state $n$ where one electron is coherently transferred from the left to the right lead (or vice versa). The stationary probabilities for the empty and occupied dot $P_0$ and $P_1$ are obtained from the rate equations
\be\label{rateeqs}
0=\frac{dP_0}{dt}=P_1 W^- - P_0 W^+,\qquad P_0+P_1=1,
\ee
where $W^\pm\equiv W^\pm_L+W^\pm_R$. It is important to note that in this case, there are no cotunneling contributions to the rate equations: Elastic cotunneling does not change the charge state of the dot, and hence does not affect the probability distribution.

\emph{Sequential-tunneling contributions.}---In the first step, we calculate the stationary probabilities and the sequential-tunneling current. The transition rates obtained from Fermi's golden rule are
\be 
W^+_a=\frac{1}{\hbar}\Gamma_a f_a(\epsilon_d), \qquad W^-_a=\frac{1}{\hbar}\Gamma_a [1-f_a(\epsilon_d)].
\ee
From substituting this into Eq.~\eqref{rateeqs} one obtains 
\begin{align}\label{prob1}
P_0&=\frac{W^-}{W^-+W^+}=\frac{\sum_a \Gamma_a [1-f_a(\epsilon_d)]}{\Gamma_L+\Gamma_R}\\\label{prob2}
P_1&=\frac{W^+}{W^-+W^+}=\frac{\sum_a \Gamma_a f_a(\epsilon_d)}{\Gamma_L+\Gamma_R},
\end{align}
and upon substituting into Eq.~\eqref{current}, the sequential-tunneling current is found to be
\begin{align}
I^\text{(seq)}&=\frac{e}{\hbar}\frac{\Gamma_L\Gamma_R}{\Gamma_L+\Gamma_R}\left[f_L(\epsilon_d)-f_R(\epsilon_d)\right]
\end{align}
Indeed, this expression is indentical with the leading-order current contribution from the expansion of the exact solution, see Eq.~\eqref{IfinTx}.

\emph{Cotunneling contributions.}---In the second step, we calculate the cotunneling current by evaluating the cotunneling rates
\be\label{cotrateT}
W^{nn}_{ab}=\frac{\Gamma_a\Gamma_b}{2\pi\hbar}\int d\epsilon\abs{\frac{1}{\epsilon-\epsilon_d}}^2f_a(\epsilon)[1-f_b(\epsilon)].
\ee
After regularization, we find
\begin{align}
&W^{nn}_{ab}=\frac{\beta}{4\pi^2}\frac{\Gamma_a\Gamma_b}{\hbar}n_B(\mu_b-\mu_a)\\\nonumber
&\times\iim \left\{\textstyle  \psi'\left(\frac{1}{2}+i\frac{\beta}{2\pi}[\mu_b-\epsilon_d]\right) - \psi'\left(\frac{1}{2}+i\frac{\beta}{2\pi}[\mu_a-\epsilon_d]\right)\right\}.
\end{align}
Substituting this together with Eqs.~\eqref{prob1}, \eqref{prob2} into the expression for the cotunneling current from Eq.~\eqref{current}, and using $n_B(x)+n_B(-x)=-1$ leads to
\begin{align}
I^\text{(cot)}=&\frac{e}{h}\Gamma_L\Gamma_R\frac{\beta}{2\pi}\iim \sum_a a  \psi'\left(\frac{1}{2}+i\frac{\beta}{2\pi}[\epsilon_d+aeV/2]\right) .
\end{align}
By comparison with Eq.~\eqref{IfinTx} one verifies that this is identical with the next-to-leading order term of the expansion of the exact solution.

\subsection{Zero-frequency noise}
Our starting point for the exact noise calculation is the two-terminal formula for the zero-frequency noise,
\begin{align}\label{blnoise}
S=\frac{2e^2}{h}\int dE\, &\bigg\{ \sum_a T(E) f_a(E)\left[1-f_a(E)\right]\\\nonumber
&+T(E)\left[ 1-T(E)\right] \left[ f_L(E) - f_R(E) \right]^2
\bigg\},
\end{align}
see Refs.\ \onlinecite{lesovik1,blanter}.
The transmission coefficient $T(E)$ is known to be proportional to the spectral function $A_d(E)$.  The exact relation between them can be extracted easily by comparing Eq.~\eqref{landauer} to the Landauer formula 
\begin{align}
\langle I \rangle &= \frac{e}{h} \int dE\, T(E) \left[ f_L(E) - f_R(E) \right].
\end{align}
This leads to $T(E) = A_d(E)\Gamma_L\Gamma_R/\Gamma$.

The integrations can be carried out by invoking the Fourier transform for the Lorentzian. The energy-integration then turns into a Fourier transform of the Fermi factors, which may be carried out by contour integration. The remaining time-integration leads to polygamma functions. The exact result for the zero-frequency noise is given by
\begin{widetext}
\begin{align}\nonumber
S=&\frac{2e^2}{h}\bigg\{
\frac{2\Gamma_L\Gamma_R(\Gamma_L^2+\Gamma_R^2)}{\Gamma^3}\coth (\beta eV/2) \iim \sum_a a  \psi\left( \frac{1}{2} +i\frac{\beta}{2\pi}[\epsilon_d+aeV/2] +\frac{\beta\Gamma}{4\pi}\right) \\\nonumber
& + \frac{2(\Gamma_L\Gamma_R)^2}{\Gamma^2}\frac{\beta}{2\pi}\coth (\beta eV/2) \iim \sum_a a \psi'\left( \frac{1}{2} +i\frac{\beta}{2\pi}[\epsilon_d+aeV/2] +\frac{\beta\Gamma}{4\pi}\right) \\
&+\frac{2(\Gamma_L\Gamma_R)^2}{\pi\Gamma^3}\rre \sum_a \psi'\left( \frac{1}{2} +i\frac{\beta}{2\pi}[\epsilon_d+aeV/2] +\frac{\beta\Gamma}{4\pi}\right) 
 -\frac{(\Gamma_L\Gamma_R)^2}{\pi\Gamma^2}\frac{\beta}{2\pi}\rre \sum_a  \psi''\left( \frac{1}{2} +i\frac{\beta}{2\pi}[\epsilon_d+aeV/2] +\frac{\beta\Gamma}{4\pi}\right)\bigg\}\label{exact}
\end{align}
Expanding this expression in $\Gamma$, we find in leading and next-to-leading order
\begin{align}\label{noiseexp1}
S^{(1)}
&=\frac{2e^2}{h}\bigg\{
\frac{2\pi \Gamma_L\Gamma_R(\Gamma_L^2+\Gamma_R^2)}{\Gamma^3}\left[ f_R(\epsilon_d)+f_L(\epsilon_d)-2f_L(\epsilon_d)f_R(\epsilon_d) \right]\\\nonumber
&\qquad \qquad
+\frac{4\pi(\Gamma_L\Gamma_R)^2}{\Gamma^3}\left( f_L(\epsilon_d)[1-f_L(\epsilon_d)] + f_L(\epsilon_d)[1-f_L(\epsilon_d)]  \right)\bigg\}\\
S^{(2)}&=\frac{2e^2\Gamma_L\Gamma_R}{h}\frac{\beta}{2\pi}\coth (\beta eV/2) \iim \sum_a a\, \psi'\left( \frac{1}{2} +i\frac{\beta}{2\pi}[\epsilon_d+aeV/2]\right) \label{noiseexp2}
\end{align}
\end{widetext}
Again, we can compare this expansion with the results from the rate-equations approach. Employing the formalism described in Appendix \ref{app:noise}, we obtain the occupation probabilities
\be
P_0=\frac{\sum_a\Gamma_a[1-f_a]}{\Gamma_L+\Gamma_R}, \qquad P_1=\frac{\sum_a\Gamma_a f_a}{\Gamma_L+\Gamma_R},
\ee
the steady-state current
\be
\langle I \rangle/  e=\left[ \frac{\Gamma_L\Gamma_R}{\Gamma_L+\Gamma_R}(f_L-f_R) + w_{LR}-w_{RL}\right],
\ee
and the vectors
\begin{align}\allowdisplaybreaks
\vc{u}_{LL}&=\left( 
\begin{array}{c}
P_0(\Gamma_L f_L/\hbar + w_{LR}+w_{RL})\\P_1(\Gamma_L[1-f_L]/\hbar+w_{LR}+w_{RL})	
\end{array}\right),\\
\vc{u}_{RR}&=\left(
\begin{array}{c}
 P_0(\Gamma_R f_R/\hbar+w_{LR}+w_{RL})\\P_1(\Gamma_R[1-f_R]/\hbar + w_{LR}+w_{RL})	
\end{array}\right),\\
\vc{u}_{RL}&=\vc{u}_{LR}=\left(
\begin{array}{c}
P_0(w_{LR}+w_{RL})\\P_1(w_{LR}+w_{RL})	
\end{array}\right),\\
\vc{y}_{L}&=\left(
\begin{array}{c}
-P_1\Gamma_L[1-f_L]/\hbar+P_0(w_{LR}-w_{RL}) \\P_0\Gamma_Lf_L/\hbar + P_1(w_{LR}-w_{RL})	
\end{array}\right),\\
\vc{y}_{R}&=\left(
\begin{array}{c}
P_1\Gamma_R[1-f_R]/\hbar+P_0(w_{LR}-w_{RL}) \\ P_0\Gamma_Rf_R/\hbar + P_1(w_{LR}-w_{RL})	
\end{array}\right),\\
\vc{w}_{L}&=\left(
\begin{array}{c}
\Gamma_Lf_L/\hbar + w_{LR}- w_{RL}\\ -\Gamma_L[1-f_L]/\hbar + w_{LR} - w_{RL}	
\end{array}\right),\\
\vc{w}_{R}&=\left(
\begin{array}{c}
-\Gamma_Rf_R/\hbar + w_{LR}- w_{RL}\\ \Gamma_R[1-f_R]/\hbar + w_{LR} - w_{RL}	
\end{array}\right).
\end{align}
A lengthy but straightforward calculation shows that the rate-equations results for the zero-frequency noise in leading and next-to-leading order are identical with Eqs.~\eqref{noiseexp1} and \eqref{noiseexp2}.

\section{Approximate evaluation of cotunneling rates\label{app:approx}}
We derive an approximate expression for the elastic cotunneling rate $W^{00}_{00;ab}$, valid for strong electron-phonon coupling $\lambda\gg1$. Our starting point is the (unregularized) rate as obtained by Fermi's golden rule, see Eq.~\eqref{cotrate1}. This expression can be greatly simplified by noting that only a few terms in the $q''$ sum give significant contributions, which is due to the behavior of Franck-Condon matrix elements for strong electron-phonon coupling. Specifically, the relevant matrix elements read
\begin{align}
&\abs{M_{0q}}^2=\frac{\lambda^{2q}e^{-\lambda^2}}{q!}\\\nonumber
& \simeq \frac{1}{\sqrt{2\pi}}\exp\left[-\lambda^2 +2q\ln\lambda + q -q\ln q -{\textstyle\frac{1}{2}}\ln q\right],
\end{align}
where we have used Sterling's approximation for the factorial in the last step.  By inspection for $\lambda\gg1$, we find that the argument of the exponenial function is strictly negative and exhibits a single maximum at $q\approx\lambda^2$. We obtain a reasonable approximation for the strong-coupling FC matrix elements by expanding the argument to second order in $q$ around the maximum $q=\lambda^2$, which leads to the Gaussian
\be
\abs{M_{0q}}^2\approx \frac{1}{\sqrt{2\pi}\lambda} \exp\left[ -\frac{(q-\lambda^2)^2}{2\lambda^2} \right].
\ee
Consequently, as a function of $q$ the (squared) FC matrix element $\abs{M_{0q}}^2$ exhibits a single peak of width $\sim\lambda$ centered at $q=\lambda^2$. Accordingly, in Eq.~\eqref{cotrate1} we only need to account for contributions from these summands. To leading order in $1/\lambda$, we can then neglect the variation of $1/q''$ within the Gaussian peak, i.e.\ setting $q''=\lambda^2$ in the energy denominator, and noting that 
$\abs{\epsilon}\ll (\epsilon_d-\lambda^2\hbar\omega)$ for bias and gate voltages small compared to the polaron shift, we obtain the approximation
\begin{align}
&W^{00}_{00;ab}\approx \frac{\Gamma_a\Gamma_b}{\pi\hbar}\frac{k_BT+eV}{(\epsilon_d+\lambda^2\hbar\omega)^2}\left(\sum_{q''} \abs{M_{0q''}}^2\right)^2\\\nonumber
&\approx\frac{\Gamma_a\Gamma_b}{\pi\hbar}\frac{k_BT+eV}{(\epsilon_d+\lambda^2\hbar\omega)^2} \left(\int_{-\infty}^\infty dq\, \frac{1}{\sqrt{2\pi}\lambda} \exp\left[ -\frac{(q-\lambda^2)^2}{2\lambda^2} \right]\right)^2\\\nonumber
&\approx \frac{\Gamma_a\Gamma_b}{\pi\hbar}\frac{k_BT+eV}{(\epsilon_d+\lambda^2\hbar\omega)^2}.
\end{align}

\section{Formalism for calculating current and shot noise\label{app:noise}}
We review Korotkov's technique for noise calculations, Ref.~\onlinecite{korotkov}, and slightly generalize his formalism in order to  incorporate higher-order processes such as cotunneling. Throughout this section, we will keep wording and formalism as general as possible to stress the wide applicability of this approach.

We consider a system consisting of some central structure, such as a quantum dot or molecule, and source and drain electrodes coupled to the central structure. We assume that the state space of the center is discrete and finite, and we denote it by
\be
\mathbb{S}\equiv \{ 1,2,\ldots,N \}.
\ee
Then, the system's dynamics originates from transitions between pairs of these states, $\ket{i}\to\ket{f}$, which are associated with specific transition rates $W_{if}\ge 0$.  Mathematically speaking, the dynamics can be classified as a time-dependent Markov process, and it is fully characterized by the Master equations
\be\label{master}
\frac{\partial}{\partial t}P(f,t|i) = \sum_{k\in\mathbb{S}}\left[ P(k,t|i) W_{kf} - P(f,t|i) W_{fk}\right]
\ee
with initial condition $P(f,t=0|i)=\delta_{if}$. Here, $P(f,t|i)$ denotes the conditional probability for the system to occupy the state $\ket{f}$ at time $t$, given that the initial state at time $t=0$ was $\ket{i}$. Note that, due to our choice of labelling processes and corresponding rates with the center degrees of freedom only, rates $W_{ii}$ with identical initial and final states exist and need to be taken into account. [A relevant example is given by elastic cotunneling processes.]

It is convenient to translate Eq.~\eqref{master} into a compact matrix notation. This can be achieved by the definitions 
\be
\vc{p}_i(t) \equiv \left( P(1,t|i),P(2,t|i),\ldots,P(N,t|i) \right)^\top,
\ee
and
\be
\mathsf{W}=\left(
\begin{array}{cccc}
-\sum_{k\not=1} W_{1k} & W_{21} & W_{31} & \cdots\\
W_{12} & -\sum_{k\not=2} W_{2k} & W_{32} & \cdots\\
W_{13} & W_{23} & -\sum_{k\not=3} W_{3k} & \cdots\\
\vdots &        &                        & \ddots 
\end{array}
\right)	,
\ee
where all $k$-sums run over the state space $\mathbb{S}$. Now, the Master equation can be summarized into
\be
\frac{\partial}{\partial t}\vc{p}_i(t)=\mathsf{W}\vc{p}_i(t)
\ee
with initial condition $\vc{p}_i(t=0)=\vc{\hat{e}}_i$, having the formal solution
\be
\vc{p}_i(t) = e^{\mathsf{W}t}\vc{\hat{e}}_i.
\ee
Under rather general conditions, typically satisfied in physical applications, one can show that for long times $t$, the probability distribution converges to a unique stationary distribution $\vc{P}$. Instead of performing the limit
\be
\vc{P}=\lim_{t\to\infty} \vc{p}_i(t) \qquad \text{(independent of $i$)},
\ee
one can solve the stationary master equation 
\be
\mathsf{W} \vc{P} = 0
\ee
with the condition that the sum over all components of the vector $\vc{P}$ is 1 (normalization), which we write as $\tr\vc{P}=1$.

\subsection{Current}
Electron transfer between the electrodes and the central system gives rise to a current, whose time-average $\langle I \rangle$ (the stationary current) must be identical in the left and right junction, i.e. $\langle I \rangle = \langle I_L \rangle = \langle I_R \rangle$, due to charge conservation. [Here, and in the following, $\langle \cdot\rangle$ always denotes time averaging.] At this point, we may fix the sign of the current once and for all, defining the current to be positive for an electron transfer from the left to the right.

Some caution is now in place regarding the nature of the transitions $\ket{i}\to\ket{f}$. In general, there may be several different \emph{processes} which cause a transition between the central system's states $\ket{i}$ and $\ket{f}$, and they differ from each other in the resulting leads' states. In order to elucidate this situation, we give one concrete example. Consider a system, whose central system is simply a single electronic level.  Then, cotunneling will result in a transition $n\to n$, i.e.\ the electronic occupation of the center remains invariant. However, there are several incoherent contributions to this transition: The cotunneling transition can result in a transfer of one electron from the left to the right, or vice versa. In principle, the tunneling back and forth between only one lead and the center is possible as well.

This example reveals that, in general, each transition $\ket{i}\to\ket{f}$ may be made up of several elementary processes contributing to this transition. Accordingly, the rates $W_{if}$ are a sum of the rates of the corresponding elementary processes $\nu$,
\be
W_{if} = \sum_\nu W_{if}^{(\nu)}.
\ee
[Of course, the notion of elementary processes only comes about due to our choice of expressing all results in terms of the states of the central system alone.]

Generally, not all elementary processes contribute to the current, and each process which does, can transfer a number $n$ of electrons either in the positive or negative direction. Taking this into account, we define the quantity $s_{if,\nu}^a$ as follows: If the elementary process $\nu$ belonging to the transition $\ket{i}\to\ket{f}$ gives a current contribution in junction $a$ ($a=L,R$), then $s_{if,\nu}^a$ gives the number of electrons transferred across junction $a$ within this process, and its sign reflects whether the transfer occurs in the positive or negative direction. For all processes without current contribution, $s_{if,\nu}^a$ vanishes. Now the stationary current in junction $a$ can be written as
\be\label{ct}
\langle I_a \rangle/e = \sum_{i,f\in\mathbb{S}}\sum_{\nu} s_{if,\nu}^a P_i W_{if}^{(\nu)} = \tr \mathsf{W}_I\vc{P},
\ee
where $(W_I)_{ij}=\sum_\nu s_{ji,\nu}^a W_{ji}^{(\nu)}$.

For the analysis of the current shot noise, we are also interested in the explicit time-dependence of the current. Viewing the system dynamics as a Markov process consisting of quasi-instantaneous jumps between different states, we can define the times $t_n^{(if,\nu)}$ ($n=1,2,\ldots,\infty$) at which the process $\nu$ belonging to the transition $\ket{i}\to\ket{f}$ takes place. Then, the time-dependent current in junction $a$ can formally be written as a sum of $\delta$ functions according to
\be
I_a(t)=e\sum_{i,f\in\mathbb{S}}\sum_{\nu,n} s_{if,\nu}^a \delta(t-t_n^{(if,\nu)}).
\ee 
Taking the time average of this expression and comparing with Eq.~\eqref{ct}, we obtain the relation
\be\label{aux}
\sum_n \langle \delta(t-t_n^{(if,\nu)}) \rangle = P_i W_{if}^{(\nu)}.
\ee

\subsection{Noise}
Our central goal consists of the derivation of a compact and general expression for the current noise defined by
\be\label{noise}
S_{ab}(\omega)=2 \int_{-\infty}^\infty d\tau\, e^{i\omega\tau}\left[ \langle I_a(\tau)I_b(0)\rangle - \langle I \rangle^2 \right].
\ee
The starting point for our considerations is the current-current correlator $\langle I_a(t)I_b(t')\rangle$. When substituting Eq.~\eqref{ct} into the correlator, we obtain a multiple sum with each summand being proportional to the average of a product of two $\delta$ functions: 
\begin{align}
\langle I_a(t)I_b(t')\rangle= &e^2 \sum_{i,f\in\mathbb{S}}\sum_{\nu,n} \sum_{i',f'\in\mathbb{S}}\sum_{\mu,m} s_{if,\nu}^a s_{i'f',\mu}^b\\\nonumber
 &\times\langle\delta(t-t_n^{(if,\nu)}) \delta(t'-t_m^{(i'f',\mu)}) \rangle.
\end{align}
 As the set of all times $t_n^{(if,\nu)}$ forms the time lattice of the Markov process, the times are considered to be pairwise distinct. Thus, we have $t_n^{(if,\nu)}\not=t_m^{(i'f',\mu)}$ for $(n,if,\nu)\not=(m,i'f',\mu)$. Accordingly, the correlator can be separated into autocorrelation and cross-correlation contributions, $\langle I_a(t)I_b(t')\rangle_a=\langle I_a(t)I_b(t')\rangle_\text{a}+\langle I_a(t)I_b(t')\rangle_\text{x}$, where
\begin{align}
\langle I_a(t)I_b(t')\rangle_\text{a}&= e^2 \sum_{i,f\in\mathbb{S}}\sum_{\nu,n} s_{if,\nu}^a s_{if,\nu}^b\\\nonumber
&\quad\times\langle \delta(t-t_n^{(if,\nu)}) \delta(t'-t_n^{(if,\nu)}) \rangle, \\
\langle I_a(t)I_b(t')\rangle_\text{x}&=\text{``non-diagonal terms"}.
\end{align}
We first turn to the evaluation of the autocorrelations. Using the fact that $\langle \delta(t-t_n^{(if,\nu)}) \delta(t'-t_n^{(if,\nu)}) \rangle=\delta(t-t')\langle \delta(t-t_n^{(if,\nu)})) \rangle$, and applying relation Eq.~\eqref{aux}, we obtain
\be
\langle I_a(t)I_b(t')\rangle_\text{a}=  e^2 \delta(t-t')\sum_{i,f\in\mathbb{S}}\sum_{\nu} s_{if,\nu}^a s_{if,\nu}^b P_i W_{if}^{(\nu)}
\ee

The cross-correlation terms now involve only distinct processes at different times. These contributions can be dealt with in the following way. Let us assume that $t\ge t'$. By using $I=dQ/dt$, we note that 
\be
\langle I_a(t) I_b(t') \rangle_\text{x} = \frac{\langle dQ_a(t) dQ_b(t') \rangle}{dt\, dt'},
\ee
and we only get a contribution to the time average, if there are current-contributing processes [causing charge variations $dQ$] in junction $b$ during $[t',t'+dt']$, and in junction $a$ during $[t,t+dt]$, which leads us to
\begin{align}
&\langle I_a(t) I_b(t') \rangle_\text{x} \\\nonumber
&\quad= e^2\sum_{i,j,k,f\in\mathbb{S}}\sum_{\mu,\nu} s_{ij,\nu}^a s_{kf,\mu}^b P_i W_{ij}^{(\nu)} p(k,t-t'|j) W_{kf}^{(\mu)}.
\end{align}
We are now in the position to give an intermediate result for the bracketed expression in Eq.~\eqref{noise}. We find
\begin{align}
&\langle \delta I_a(\tau)\delta I_b(0)\rangle = e^2 \delta(\tau)\sum_{i,f\in\mathbb{S}}\sum_{\nu} s_{if,\nu}^a s_{if,\nu}^b P_i W_{if}^{(\nu)}\\\nonumber
&\quad+e^2\sum_{i,j,k,f\in\mathbb{S}}\sum_{\mu,\nu} s_{ij,\nu}^a s_{kf,\mu}^b P_i W_{ij}^{(\nu)} \left[p(k,\tau|j)-P_k\right] W_{kf}^{(\mu)}.
\end{align}
This expression is valid for $\tau\ge 0$. For $\tau<0$ we exploit the symmetry
\be
\langle \delta I_a(\tau)\delta I_b(0)\rangle=\langle \delta I_a(0)\delta I_b(-\tau)\rangle.
\ee
Now, the last missing piece for an evaluation of the noise consists of the determination of the Fourier transform of the term
\be\label{gkj}
G_{kj}(\tau)=\theta(
\tau)\left[p(k,\tau|j)-P_k\right].
\ee
 Noting that $p(k,\tau|j)=(e^{\mathsf{W}\tau})_{kj}$ and defining the matrix $\mathsf{V}$ by $V_{kj}=P_k$, we can rewrite Eq.~\eqref{gkj} in matrix form as $\mathsf{G}(\tau)=\theta(\tau)\left[e^{\mathsf{W}t}-\mathsf{V}\right]$, and its Fourier transform is found to be
\be
\mathsf{G}(\omega)=\int_0^\infty dt\, e^{i\omega \tau}\left[ e^{\mathsf{W} \tau}-\mathsf{V}\right] = -(\mathsf{W}+i\omega\openone)^{-1} +\frac{1}{i\omega}\mathsf{V},
\ee
which can directly be evaluated at all finite frequencies. (The $\omega\to0$ will be discussed below.)
Thus, the noise can be expressed as
\begin{align}\label{sab}
&S_{ab}(\omega)=2e^2\sum_{i,f\in\mathbb{S}}\sum_{\nu} s_{if,\nu}^a s_{if,\nu}^b P_i W_{if}^{(\nu)}\\\nonumber
&\quad+\bigg[2e^2 \sum_{i,j,k,f\in\mathbb{S}}\sum_{\mu,\nu} s_{ij,\nu}^a s_{kf,\mu}^b P_i W_{ij}^{(\nu)} G_{kj}(\omega) W_{kf}^{(\mu)}
\\\nonumber
&\qquad\qquad
+ (\text{same term with } a\leftrightarrow b)^* \bigg]\\\nonumber
&\qquad=2e^2\left[ \tr \vc{u}_{ab} + \vc{w}_b\mathsf{G}(\omega)\vc{y}_a +\vc{w}_a\mathsf{G}(\omega)^*\vc{y}_b \right].	
\end{align}
Here we have defined the vectors $\vc{u}_{ab}$, $\vc{y}_a$ and $\vc{w}_b$ by
\begin{align}
(u_{ab})_i &= \sum_{f\in\mathbb{S}} \sum_\nu s_{if,\nu}^a s_{if,\nu}^b P_i W_{if}^{(\nu)}\\
(y_a)_j    &= \sum_{i\in\mathbb{S}} \sum_\nu s_{ij,\nu}^a P_i W_{ij}^{(\nu)}\\
(w_b)_k    &= \sum_{f\in\mathbb{S}} \sum_\mu s_{kf,\mu}^b W_{kf}^{(\mu)}
\end{align}

In order to evaluate the zero-frequency limit $\omega\to0$, we make use of the following properties of the relevant matrices. First, we observe that $\mathsf{G}\vc{P}=0$. Second, one easily confirms that $\mathsf{V}\vc{x}=0$ for any traceless vector $\vc{x}$. Third, we note that any well-behaved rate-equations matrix $\mathsf{W}$ (which has a unique steady-state solution) is invertible in the subspace of traceless vectors. Thus, in Eq.~\eqref{sab} we may modify the vector $\vc{y}_a$ by adding some multiple of the stationary distribution $\vc{P}$, leaving the result invariant.

We choose
\begin{align}
(\bar{y}_a)_j&\equiv (y_a)_j -\langle I_a \rangle P_j/e,
\end{align} 
which makes $\bar{\vc{y}}_a$ a traceless vector. Consequently, the zero-frequency noise can be evaluated via
\begin{align}
S_{ab}(\omega=0)&=2e^2\left[ \tr \vc{u}_{ab} - \vc{w}_b\mathsf{W}^{-1}\bar{\vc{y}}_a -\vc{w}_a\mathsf{W}^{-1}\bar{\vc{y}}_b \right].	
\end{align}

\section{Zero-frequency noise and Fano factor for the telegraph regime\label{app:telegraph}}
According to the simplest situation where telegraph noise occurs, we take into account the transitions (1) $\ket{0,0}\to\ket{1,0}$ and $\ket{1,0}\to\ket{0,0}$ due to sequential tunneling, and (2) $\ket{0,0}\to\ket{0,0}$ and $\ket{1,1}\to\ket{1,1}$ due to cotunneling. At finite bias the transport is essentially unidirectional, so that we restrict to $LR$ cotunneling, as well as sequential tunneling $\ket{0,0}\to\ket{1,0}$ via the left junction and $\ket{1,0}\to\ket{0,0}$ via the right junction. In order to avoid irrelevant indices, we will here denote the rates as $W^+$, $W^-$ for sequential tunneling, and $W^0$, $W^1$ for cotunneling. The rate-equation matrix then reads
\be
\mathsf{W}=\left(
\begin{array}{rr}
-W^+ & W^-\\
W^-  & -W^+
\end{array}\right),
\ee
and the resulting steady-state occupation probabilities for the empty and singly-occupied molecule are
\be
P_0=\frac{W^-}{W^+ + W^-}, \quad P_1=\frac{W^+}{W^+ + W^-}.
\ee
Note that due to the spin-degeneracy of the singly-occupied state, the sequential-tunneling rates obey $W^+=2W^-$, leading to $P_0=1/3$ and $P_1=2/3$. Neglecting sequential-tunneling rates as compared to cotunneling rates where appropriate, we find that the vectors $\vc{u}_{aa'}$, $\vc{y}_a$, and $\vc{w}_a$ are given by
\begin{align}
\vc{u}_{aa'} &= { P_0 W^0 \choose P_1 W^1} = \vc{y}_a,\\
\vc{w}_a     &= { W^0 \choose W^1}.
\end{align}
The steady-state current is 
\be
\langle I \rangle = P_0 W^0 + P_1 W^1=\frac{1}{3}W^0+\frac{2}{3}W^1,
\ee
so that $\bar{\vc{y}}_a$ is obtained as
\be
\bar{\vc{y}}_L=\bar{\vc{y}}_R=\frac{W^+W^-(W^0-W^1)}{(W^+ + W^-)^2}{1 \choose -1}.
\ee
Finally, we have
\be
\mathsf{W}^{-1}\bar{\vc{y}}_a=- \frac{W^+W^-(W^0-W^1)}{(W^+ + W^-)^3}{1 \choose -1}.
\ee
For the noise, we therefore find $S_{LL}=S_{RR}=S_{LR}=S_{RL}$, and the final results are
\begin{align}
S(\omega=0)&\approx 4e^2\frac{W^+W^-(W^0-W^1)^2}{(W^++W^-)^3},\\
F&\approx 6\frac{W^+W^-(W^0-W^1)^2}{(W^0+2W^1)(W^++W^-)^3}.
\end{align}

\end{document}